\documentclass[twocolumn,superscriptaddress,secnumarabic,amssymb, nobibnotes, aps, linen]{revtex4-2}
\usepackage{graphicx}
\usepackage{bm}
\usepackage{xcolor}
\usepackage{amsmath}
\usepackage{ulem}
\usepackage{comment}
\definecolor{CNRSIntermediateBlue}{RGB}{83,148,184}

\begin{document}
\title{Valence can control the nonexponential viscoelastic relaxation of multivalent reversible gels}
\author{Hugo \surname{Le Roy} $^\dag$}
\email{h.leroy@epfl.ch}
\affiliation{Université Paris-Saclay, CNRS, LPTMS, 91405, Orsay, France}
\affiliation{Institute of Physics, \'Ecole Polytechnique F\'ed\'erale de Lausanne (EPFL), 1015 Lausanne, Switzerland}
\author{Jake Song}
\email{These authors contributed equally to this work}
\affiliation{Department of Materials Science and Engineering, Massachusetts Institute of Technology, 77 Massachusetts Avenue, Cambridge, MA 02139, USA}
\author{David Lundberg}
\affiliation{Department of Chemical Engineering, Massachusetts Institute of Technology, 77 Massachusetts Avenue, Cambridge, MA 02139, USA}
\author{Aleksandr V. Zhukhovitskiy}
\affiliation{Department of Chemistry, Massachusetts Institute of Technology, 77 Massachusetts Avenue, Cambridge, MA 02139, USA}
\affiliation{Department of Chemistry, University of North Carolina at Chapel Hill; Chapel Hill, NC 27599, USA}
\author{Jeremiah A. Johnson}
\affiliation{Department of Chemistry, Massachusetts Institute of Technology, 77 Massachusetts Avenue, Cambridge, MA 02139, USA}
\author{Gareth H. McKinley}
\affiliation{Department of Mechanical Engineering, Massachusetts Institute of Technology, 77 Massachusetts Avenue, Cambridge, MA 02139, USA}
\author{Niels Holten-Andersen}
\affiliation{Department of Materials Science and Engineering, Massachusetts Institute of Technology, 77 Massachusetts Avenue, Cambridge, MA 02139, USA}
\author{Martin Lenz}
\email{martin.lenz@universite-paris-saclay.fr}
\affiliation{Université Paris-Saclay, CNRS, LPTMS, 91405, Orsay, France}
\affiliation{PMMH, CNRS, ESPCI Paris, PSL University, Sorbonne Université, Université de Paris, F-75005, Paris, France}
\begin{abstract}
Gels made of telechelic polymers connected by reversible crosslinkers are a versatile design platform for biocompatible viscoelastic materials. Their linear response to a step strain displays a fast, near-exponential relaxation when using low valence crosslinkers, while larger supramolecular crosslinkers bring about much slower dynamics involving a wide distribution of time scales whose physical origin is still debated. 
Here, we propose a model where the relaxation of polymer gels in the dilute regime originates from elementary events in which the bonds connecting two neighboring crosslinkers all disconnect. Larger crosslinkers allow for a greater average number of bonds connecting them, but also generate more heterogeneity. 
We characterize the resulting distribution of relaxation time scales analytically, and accurately reproduce stress relaxation measurements on metal-coordinated hydrogels with a variety of crosslinker sizes including ions, metal-organic cages, and nanoparticles. 
Our approach is simple enough to be extended to any crosslinker size and could thus be harnessed for the rational design of complex viscoelastic materials.
\end{abstract}
\maketitle




\section*{Introduction}
Soft hydrogels are ubiquitous in biology and dictate the mechanics of cells and tissues~\cite{Ronsin_2016}. Due to their biocompatibility, synthetic hydrogels are thus prime candidates to serve as robust soft tissue implants, although fine control of their viscoelastic properties is crucial for their success in this role~\cite{Ana_Acta_2011,chaudhuri2015substrate}. In simple viscoelastic materials, stress relaxes according to a single exponential with a single relaxation time. This is not however the case for most biological materials such as cells~\cite{kollmannsberger2011linear}, tissues~\cite{chaudhuri2020effects}, mucus~\cite{celli2009helicobacter}, and biofilms~\cite{fabbri2016mechanical}. Instead, their relaxation is characterized by a broad distribution of relaxation times~\cite{song2022non}. Such relaxation is often heuristically described by a stretched exponential:
\begin{equation}\label{eq1}
\sigma(t)\propto e^{-(t/\tau)^{\alpha}}, 
\end{equation}
where smaller values of the exponent $\alpha \in ]0;1[$ denote broader distributions of relaxation time scales~\cite{Bouchaud_stretch_exponential}.
Other similarly phenomenological fitting functions include power-law dependences of $\sigma$ on $t$~\cite{scott1959creep, keshavarz2017nonlinear, ballandPowerLawsMicrorheology2006} and log-normal distributions of the relaxation times~\cite{epstein2019modulating,masurel2015role}.

\begin{figure}[t]
\centering
\includegraphics[height=7.48cm]{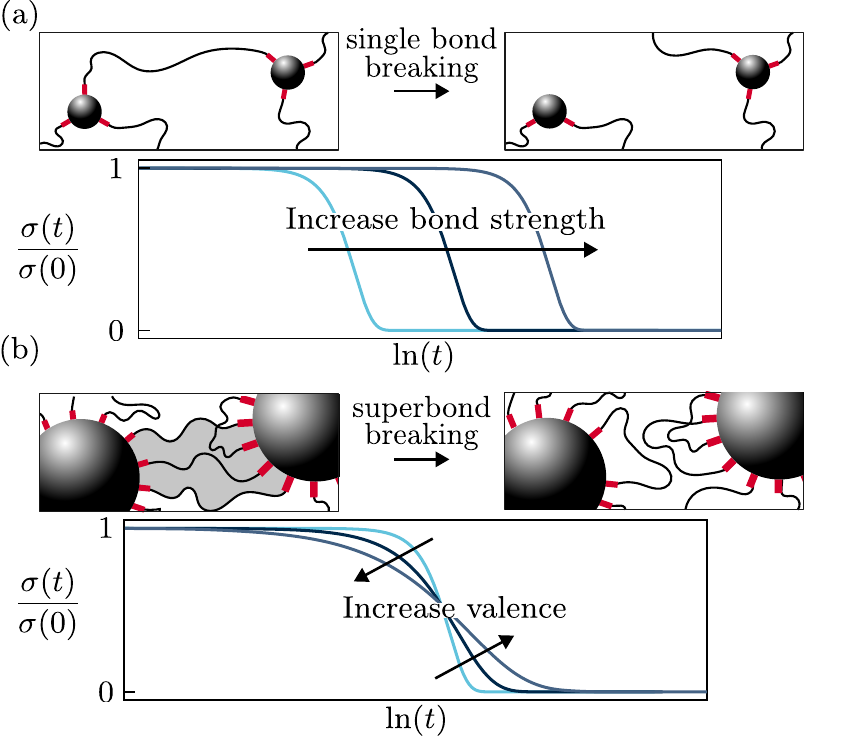}
\caption{\label{crosslink} High-valence crosslinkers yield a slow, potentially complex unbinding dynamics 
(a)~Hydrogels held together by small crosslinkers relax over the time scale associated with the unbinding of a single polymer strand. 
There, stress decays exponentially in response to a step strain: $\sigma(t)/\sigma(0)=\exp(-t/\tau)$. Since $t/\tau=\exp(\ln t-\ln\tau)$, changes in the time scale $\tau$ shift the $\sigma$ \textit{vs.} $\ln t$ relaxation curve horizontally, but do not alter its shape.
(b)~In contrast, relaxation events in the presence of high-valence crosslinkers require the simultaneous unbinding of many polymer stands. 
The associated time scale is long and highly variable depending on the number of strands involved in the ``superbond'' (grey shade). 
As a result the stress relaxation of such gels is no longer exponential and the precise shape of the relaxation curve strongly depends on the valence of the crosslinkers.}
\end{figure}

Associative gels, which relax by a succession of binding and re-binding events~\cite{parada2018ideal}, offer a promising route to design controllable viscoelastic materials. It is thus possible to tune their relaxation time by adjusting the chemical binding energy of their crosslinkers~\cite{Grindy_Nat_Mat_2016,rosales2016design}.
Although this chemistry-based approach allows tuning of the overall stress relaxation \textit{time} as illustrated in Fig.~\ref{crosslink}(a), less is known about the different approaches to tune the \textit{shape} of the stress relaxation curve of reversible hydrogels.
Accordingly, most existing models for the relaxation of multivalent gels focus on regimes dominated by a single relaxation time scale~\cite{semenovAssociatingPolymersEquilibrium1995}, leading to exponential relaxation~\cite{michelUnstableFlowNonmonotonic2001}.
Control over the distribution of relaxation time scales could however be achieved in synthetic hydrogels connected with multivalent crosslinkers such as nanoparticles~\cite{li2016controlling}, metal-organic cages~\cite{zhukhovitskiyPolymerStructureDependent2016}, clay~\cite{wang2010high}, and latex beads~ \cite{Chatterjee_MacroMol_2014}, which are known to exhibit nonexponential viscoelastic relaxation.
Here, we aim to elucidate this emergence of a wide distribution of time scales in materials with high valence crosslinkers to enable the rational design of complex gels [Fig.~\ref{crosslink}(b)]. 
Here we use the term ``valence'' to designate the number of polymer strands that a crosslinker can bind, a property sometimes also sometimes referred to as their ``functionality''~\cite{flory1953principles}. 

We propose that the emergence of a broad distribution of relaxation time scales arises from microscopic events consisting of the severing of the physical connection between two crosslinkers. 
We first propose a model where this connection, hereafter termed ``superbond'', breaks if all its constitutive crosslinkers are detached at the same time [Fig.~\ref{crosslink}(b)].
We show that the breaking time of a superbond increases exponentially with the number of strands involved, consistent with previous observation~\cite{gomez2011probing}. As a result of this strong dependence, small spatial heterogeneities in the polymer concentration may result in widely different relaxation times from one superbond to the next. Such exponential amplification of relaxation times originating from small structural differences forms the basis of models previously used to describe the relaxation of soft glasses \cite{bertinSubdiffusionLocalizationOnedimensional2003, SGR, trachenkoSlowStretchedexponentialFast2021}.
In contrast with these studies, our approach explicitly models the microscopic basis of this amplification. That allows it to not only recover relaxation curves virtually indistinguishable from those discussed in previous studies, but to also predict the influence of temperature and crosslinker valence on the macroscopic stress relaxation observed in the resulting gel.
The details of the polymer strand morphology are not central to this influence, and we thus enclose them in a few effective parameters which could be derived from first principles in specialized models related to specific implementations of our basic mechanism.
To confirm these predictions, we conduct experiments on hydrogels with four distinct crosslinker types of different sizes, and find that our model quantitatively reproduces multiple relaxation curves using this small set of microscopic parameters.
Finally, we show that several phenomenological fitting functions used in the literature can be recovered as asymptotic regimes of our analytical model.

\section*{Results}
\subsection*{Model of a single superbond}
We first model a single superbond in the simple, experimentally relevant \cite{zhukhovitskiyPolymerStructureDependent2016} case of strong interactions combined with short polymers, which implies negligible entanglements.  In the limit of very large and rigid crosslinkers, the polymer layer around a crosslinker is locally planar, its structure is not affected by small fluctuations of polymer concentration and its thickness fixes the distance between crosslinkers \cite{raspaudTriblockCopolymersSelective1996,zhukhovitskiyPolymerStructureDependent2016}. We thus use the simplifying assumption that all individual bonds participating in a superbond are identical and non-interacting, an approximation whose validity we ultimately assess through comparisons with experiments.

\begin{figure}[t]
\centering
\includegraphics[width=8.5cm]{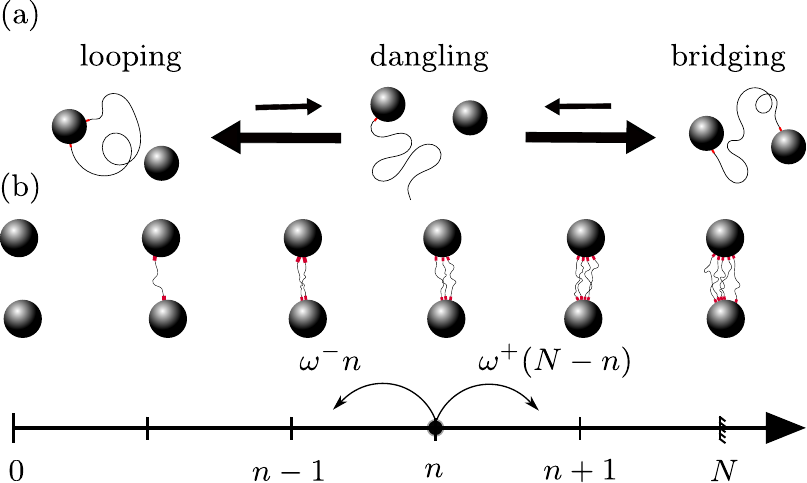}
\caption{We model superbond breaking as the disconnection of many independent polymer strands. (a)~Disconnecting a single polymer strand requires going through a high-energy, short-lived dangling state (larger arrows indicate faster transitions). The looping and bridging states both have two polymer-crosslinker bonds, and therefore have the same energy. (b)~Individual strands in a superbond attach and detach independently, resulting in a one-dimensional random walk in the number $n$ of attached strands [Eq.~\eqref{eq4}].
Here we only draw the bridging strands, not the looping strands.
}
\label{Intermediate}
\end{figure}

We model the attachment and detachment of a single polymer strand from a pair of crosslinkers as shown in Fig.~\ref{Intermediate}(a). When both its ends are bound, the strand may or may not connect two different crosslinkers. The corresponding ``bridging'' and ``looping'' states have the same energy since we assume the polymer strand to be completely flexible, and we denote by $\Delta S$ the entropy difference between them. To transition between these two states, the strand must disconnect one of its ends and form a ``dangling'' state. The disconnection of the strand in this state implies an energy barrier $\Delta E$ that is much larger than the thermal energy $k_BT=\beta^{-1}$. This implies that the dangling state is short-lived, and thus need not be explicitly included in our modeling. Our approximation scheme implies that the overall rate $\omega^+$ and $\omega^-$ to go from the looping to the bridging state and back are constant. They read
\begin{equation}
\omega^+ = \frac{1}{\tau_0} e^{-\beta \Delta E}
\qquad
\omega^- = \frac{1}{\tau_0} e^{-\beta \Delta E + \Delta S},
\label{eq2}
\end{equation}
where the typical time scale $\tau_0$ takes into account the entropy difference between the looping and dangling state. At equilibrium, we denote the probability for a single polymer strand to create a bridge as $p_\text{on}=1-p_\text{off}=1/(1+e^{\Delta S})$.

We now consider the dynamics of a superbond involving $N$ polymer strands. Within our approximation of independent attachment and detachment of the individual polymer strands, the superbond undergoes the Markov process illustrated in Fig.~\ref{Intermediate}(b) and the probability $P_n(t)$ for $n$ strands to create bridges between the two crosslinkers at time $t$ satisfies the master equation

\begin{align}
\partial_t P_n(t) = & \,(N-n+1)\,\omega^+P_{n-1}(t) + (n+1)\,\omega^-P_{n+1}(t)\nonumber\\
&-[(N-n)\,\omega^++n\,\omega^-]P_n(t), \label{eq4}
\end{align}
which ensures that the number of bridging polymers can never be greater than $N$.

To determine the rate at which a superbond breaks, we set an absorbing boundary condition $P_0(t)=0$ and define its survival probability as $S(t) = \sum_{n=1}^N P_n(t)$. In the limit $N\gg1$ where a large number of strands are involved in the superbond, we show in Sec.~S1 of the Supplementary Information that the detachment of the two beads is analogous to a Kramers escape problem. We thus prove that the survival probability decays as a single exponential $S(t) = \exp(-t/\tau_N)$~\cite{texierIndividualEnergyLevel2000} with an average detachment time
\begin{equation}\label{eq:tau_N}
\tau_N \underset{N\rightarrow\infty}{\sim} \frac{\tau_0 e^{\beta \Delta E}}{N p_\text{off}^N}.
\end{equation}
The breaking of the superbond can thus be assimilated to a Poisson process with rate $1/\tau_N$ regardless of the initial condition $P_n(0)$. The strong, exponential dependence of $\tau_N$ on $N$ implies that any dispersity in the number of strands involved in a superbond may result in a wide distribution of time scales.

\subsection*{Model of the relaxation of a gel}
Two factors influence the dispersity of $N$. First, its value is constrained by the available space at the surface of each crosslinker, which we model by setting an upper bound $N_\text{sat}$ on the number of polymer strands (in a loop or a bridge) participating in any superbond. Second, depending on the local density of polymer in the vicinity of the superbond, the actual number of strands present may be lower than $N_\text{sat}$. In the regime where the polymer solution surrounding the crosslinkers is dilute, polymer strands are independently distributed throughout the system. As a result, the distribution of local strand concentrations within a small volume surrounding a superbond follows a Poisson distribution. We thus assume that $N$ is also described by a Poisson distribution up to its saturation at $N_\text{sat}$:

\begin{equation}\label{eq10}
 p(N) = 
\begin{cases}
 \frac{\bar{N}^N e^{-\bar{N}}}{N!} &\text{for } N <N_\text{sat} \\
 \sum_{K=N_\text{sat}}^{+\infty} \frac{\bar{N}^K e^{-\bar{N}}}{K!} &\text{for } N = N_\text{sat}
 \end{cases}
,
\end{equation}
where $\bar{N}$ would be the average number of strands in a superbond in the absence of saturation and thus depends on the ratio of polymer to crosslinker concentration. Note that the specific form of the distribution used in Eq.~\eqref{eq10} does not significantly modify our results, as discussed later.

\begin{figure}[t]
    \includegraphics[width=9cm]{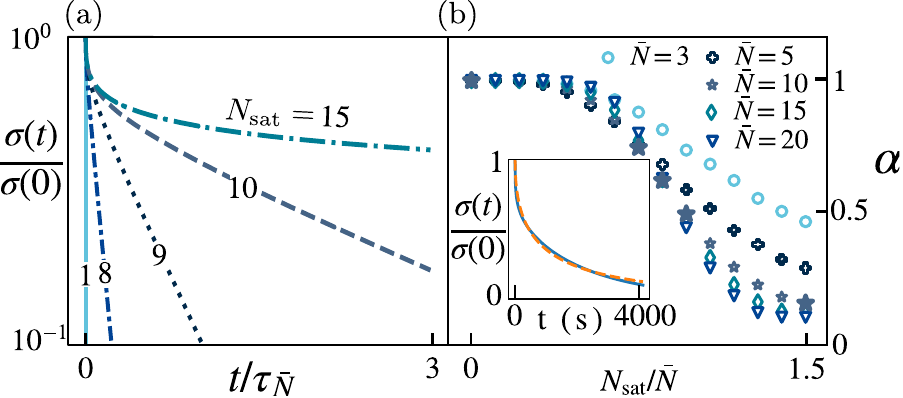}
    \caption{(a) Disperse, high-valence superbonds initially display a nonexponential mechanical relaxation, then cross over to an exponential regime when only the saturated superbonds remain. Curves plotted from Eq.~\eqref{eq:response} with $p_\text{off} = 0.2$, $\bar{N}=10$ and different values of $N_\text{sat}$ as indicated on each curve.
    (b) Relationship between the stretch exponent $\alpha$ quantifying the nonexponential character of the relaxation and the microscopic parameter $N_\text{sat}/\bar{N}$. Here $p_\text{off} = 0.2$. A low $N_\text{sat}/\bar{N}$ gives an exponential relaxation ($\alpha \simeq 1$), while a larger $N_\text{sat}/\bar{N}$ leads to a more complex behavior ($\alpha < 1$). While $\alpha$ appears to converge to a finite value for large $N_\text{sat}/\bar{N}$ for the largest values of $\bar{N}$, this behavior is contingent on our choice of fitting interval. This issue does not affect the rest of the curves. Large stars correspond to the curves represented in (a). Inset: illustration of the quality of the fits between the heuristic stretched exponential [dashed orange line, Eq.~\eqref{eq1}] and our prediction [solid blue line, Eq.~\eqref{eq:response}].} 
    \label{fig:stretched}
\end{figure}

In response to a step strain, we assume that each superbond is stretched by an equal amount and resists the deformation with an equal force prior to breaking. Superbonds may subsequently reform, but the newly formed bonds are not preferentially stretched in the direction of the step strain and therefore do not contribute to the macroscopic stress on average. Denoting by $t=0$ the time at which the step strain is applied and by $\sigma(t)$ the resulting time-dependent shear stress, the progressive breaking of the initial superbonds results in the following stress response function:
\begin{equation}\label{eq:response}
\frac{\sigma(t)}{\sigma(t=0)}=\sum_{N=1}^{N_\text{sat}} \frac{p(N)}{1-p(0)} e^{-t/\tau_N}.
\end{equation}
While the breaking times $\tau_N$ are unaffected by the applied stress in the linear response regime, nonlinearities could easily be included in our formalism by making $\Delta S$ stress-dependent and thus favor strand detachment. The relaxation described in Eq.~\eqref{eq:response} occurs in two stages. At long times $t\gg\tau_{ N_\text{sat}}$, few short-lived superbonds remain. Saturated superbonds ($N=N_\text{sat}$) dominate the response and Eq.~\eqref{eq:response} is dominated by the last term of its sum. As a result, the stress relaxes exponentially over time, as seen from the linearity of the log-lin curves of Fig.~\ref{fig:stretched}(a) for large values of $t$. Systems with smaller values of $N_\text{sat}$ manifest this regime at earlier times; in the most extreme case, the relaxation of a system where superbonds involve at most a single polymer strand ($N_\text{sat}=1$) is fully exponential and extremely fast as compared to systems with higher $N_\text{sat}$. Over short times ($t\ll\tau_{ N_\text{sat}}$), stress relaxation involves multiple time scales. This nonexponential regime is apparent on the left of Fig.~\ref{fig:stretched}(a). These two regimes have already been reported in several experimental gels connected by multivalent crosslinkers \cite{Ronsin_2016,raspaudTriblockCopolymersSelective1996}.

\begin{figure*}[t]
\includegraphics[width=17cm]{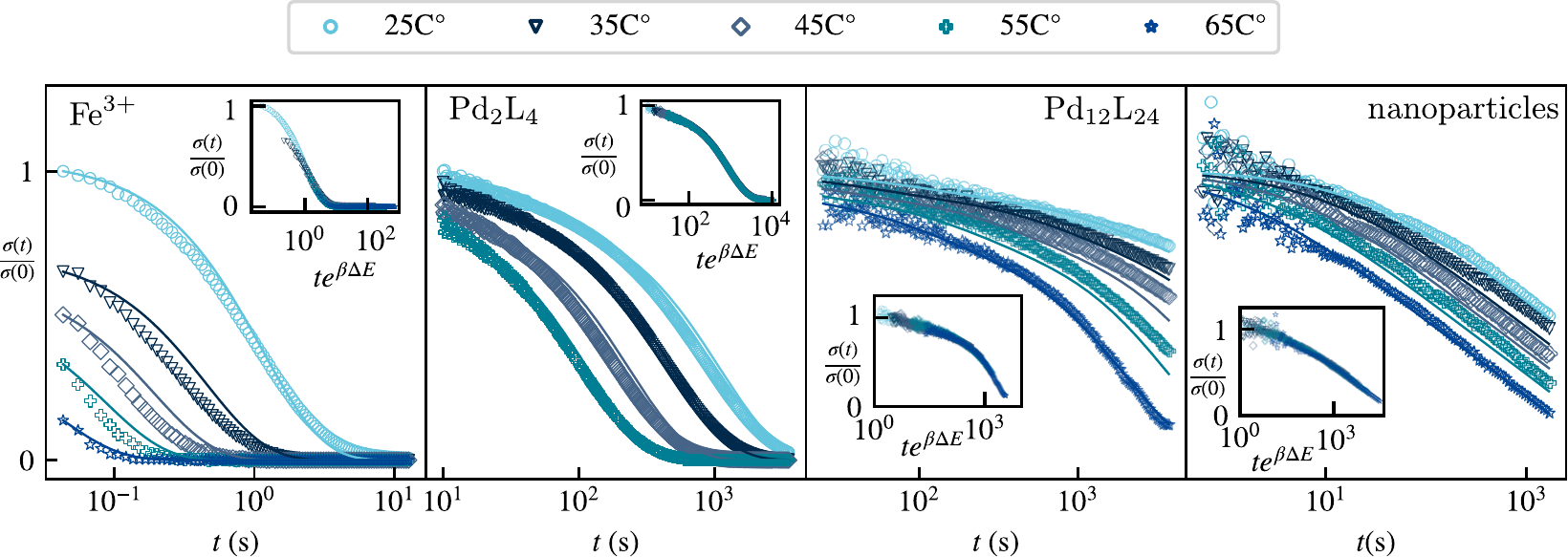}
\caption{Stress relaxation function for four experimental systems with increasing crosslinker valences (see Table~\ref{table:parameter} for values). Here we use a log-lin scale [unlike in Fig.~\ref{fig:stretched}(a)] to facilitate the visualization a large range of time scales. Alternate representations are available as Fig.~S5.
Symbols are experimental datapoints, and the lines are the associated fitting curves. 
Insets: time-temperature collapsed data obtained by a rescaling $t \rightarrow te^{\beta \Delta E}$.
}
\label{fig2}
\end{figure*}

While Eq.~\eqref{eq:response} is not identical to the stretched exponential of Eq.~\eqref{eq1}, the inset of Fig.~\ref{fig:stretched}(b) shows that they are remarkably close in practice. We thus relate the stretch exponent $\alpha$ to the saturation number $N_\text{sat}$ by fitting a stretched exponential to our predicted stress response function over the time interval required to relax $90\%$ of the initial stress [Fig.~\ref{fig:stretched}(b)]. The fits are very close matches, and consistently give correlation factors $r^2 > 0.98$ (see detailed plots in Supplementary Fig.~S2). If $N_\text{sat}\lesssim 0.5\bar{N}$ then $\alpha\simeq 1$, indicating a nearly-exponential relaxation. Indeed, in that case superbond saturation occurs well before the peak of the Poisson distribution of $N$. Physically, this implies that the local polymer concentration surrounding most superbonds is sufficient to saturate them. As almost all superbonds are saturated, they decay over the same time scale $\tau_{N_\text{sat}}$. As a result, the material as a whole displays an exponential relaxation. For larger values of $N_\text{sat}$, the Poisson distribution is less affected by the saturation and the dynamics is set by the successive decay of superbonds involving an increasing number of strands, implying lower values of $\alpha$. The larger the value of $\bar{N}$, the sharper the crossover between these two regimes.

\subsection*{Experiments}

To validate our model of the effect of crosslinker valence on hydrogel relaxation, we perform 
step-strain experiments of poly(ethylene glycol) (PEG)-based gels involving a range of crosslink valences and compare Eq.~\eqref{eq:response} to the resulting relaxation curves. We implement strand-crosslink bonds based on two sets of metal-cooordination chemistry: nitrocatechol-$\text{Fe}^{3+}$ coordination and bispyridine-$\text{Pd}^{2+}$ coordination. The first set of gels is made with nitrocatechol-functionalized PEG crosslinked by single $\text{Fe}^{3+}$ ions with an estimated valence of 3, and by iron oxide nanoparticles. The nanoparticles have a mean diameter of $7\,\text{nm}$ with a surface area that allows a valence of $\sim 100$ ligands ~\cite{li2016controlling}. The second set of gels are made with bispyridine-functionalized PEG, wherein bis-$meta$-pyridine ligands induce self-assembly of gels that are crosslinked by $\text{Pd}_{2}$$\text{L}_{4}$ nanocages with a valence of 4, and bis-$para$-pyridine ligands induce self-assembly of gels that are crosslinked by $\text{Pd}_{12}$$\text{L}_{24}$ nanocages with a valence of 24~\cite{PolyMOC}.
As shown in Fig.~\ref{fig2}, these four distinct gel designs result in a broad range of relaxation behaviors. Overall, large-valence gels and lower temperatures result in longer relaxation times, consistent with the illustration of Fig.~\ref{crosslink}(a). The relaxation curves associated to high-valence crosslinkers are also less steep, consistent with the involvement of a broader distributions of relaxation times and the schematic of Fig.~\ref{crosslink}(b).

\begin{table*}[t]
\centering
\begingroup
\setlength{\tabcolsep}{6pt} 
\renewcommand{\arraystretch}{1.2}
\begin{tabular}{|c|c|c|c|c|}
\hline
\textbf{crosslinker}  & $\text{Fe}^{3+}$& Pd$_{2}$L$_{4}$ & Pd$_{12}$L$_{24}$ & nanoparticles \\
\hline
\textbf{estimated valence}& $3$ & $4$ & $24$ & $100$\\
\hline
$\bm{N_\text{sat}}$& $1$ & $4$ & $7$ & $17$\\
\hline
$\bm{\Delta E}$ (units of $k_BT$)  & $28$ & $24$ & $24$ & $24$  \\
\hline
$\bm{p}_\mathbf{off}$ & $0.08$ & $0.15$ & $0.15$ & $0.36$ \\
\hline
$\bm{\tau_1} \text{ at $T=300\,\text{K}$}$ (s) & $1.7$ & $6.1$ & $1.9$ & $0.1$\\ 
\hline
$\bm{\bar{N}}$& $1$ & $6$ & $9$ & $14$\\
\hline
\end{tabular}
\endgroup
\caption{Estimated and fitted parameters involved in the comparison between experiment and theory in Fig.~\ref{fig2}. The energies are given in units of $k_BT$ for $T=300\,\text{K}$. Instead of displaying the parameter $\tau_0$, we present the more easily interpreted unbinding time of a single polymer strand at $300K$, namely $\tau_1 = \tau_0 e^{\beta_{300} \Delta E}/p_\text{off}$.
\label{table:parameter}}
\end{table*}

At a more detailed level, our assumption that the dynamics of single polymer strand proceeds independently of its environment implies the existence of a single energy scale $\Delta E$. As a result, we predict that all time scales involved in the relaxation are proportional to $\exp(-\beta\Delta E)$. We confirm this through a time-temperature collapse shown in the insets of Fig.~\ref{fig2} (see Sec.~S4 of the Supplementary Information for details). This collapse provides us with the value of $\Delta E$ for each of our four systems, which we report in Table~\ref{table:parameter}. The binding energy value $\Delta E$ for the $\text{Pd}_{2}$$\text{L}_{4}$ and $\text{Pd}_{12}$$\text{L}_{24}$ gels match, as expected from the fact that they originate from the same composition. On the other hand, the $\Delta E$ of the nanoparticles gel cannot be directly compared to the $\text{Fe}^{3+}$ ion gel because of their dramatic physico-chemical properties as well as the acidic pH used to facilitate mono-coordination in nanoparticles.


To compare the temperature-collapsed curves to our prediction of Eq.~\eqref{eq:response}, we fit the parameters $p_\text{off}$, $\tau_0$, $\bar{N}$ and $N_\text{sat}$ across multiple temperatures. The resulting fits, shown in Fig.~\ref{fig2}, display a good agreement between the theory and experiments across up to 4 orders of magnitude in time scales. The fitted values of $N_\text{sat}$ are moreover consistent with the estimated valence of the crosslinkers (Table~\ref{table:parameter}), which confirms our interpretation of the physical origin of $N_\text{sat}$. 
A more direct relation between the valence and $N_\text{sat}$ would require additional information about the structure of each category of gels, and in particular the number of nearest neighbors of an individual crosslink, which remains debated for these types of gels \cite{hsiaoRoleIsostaticityLoadbearing2012,colomboSelfassemblyCooperativeDynamics2014,gu2018photoswitching}. The possible clustering of crosslinkers may also influence this relation, as nanocage systems similar to ours \cite{liSpontaneousSelfAssemblyMetal2008} may form higher-order nanocage structures. 
While such complexities as well as possible imperfections in our fitting procedure complicate the literal interpretation of the fitted values of $N_\text{sat}$, these nonetheless clearly increase with increasing crosslinker valence.
The fit also supports the notion that the mean number of strands per superbond $\bar{N}$ accounts for the distribution of relaxation time scales in our gels. 
The $\text{Fe}^{3+}$ gel thus displays an exponential relaxation consistent with $\bar{N}=N_\text{sat}=1$. The higher-valence Pd$_2$L$_4$ and Pd$_{12}$L$_{24}$ systems have a complex relaxation at early times followed by an exponential behavior, as expected for $\bar{N}\simeq N_\text{sat}>1$. As expected from our model, the crossover time $\tau_{N_\text{sat}}$ separating the two regimes is larger in the higher-valence Pd$_{12}$L$_{24}$ gel. Finally, the high-valence nanoparticle system shows an extended complex relaxation associated with $\bar{N}<N_\text{sat}$, thus confirming that all the qualitative relaxation regimes discussed in the previous sections are experimentally relevant.

\subsection*{Distribution of relaxation time scales}
\begin{figure}[t]
    \centering
    \includegraphics[width=8cm]{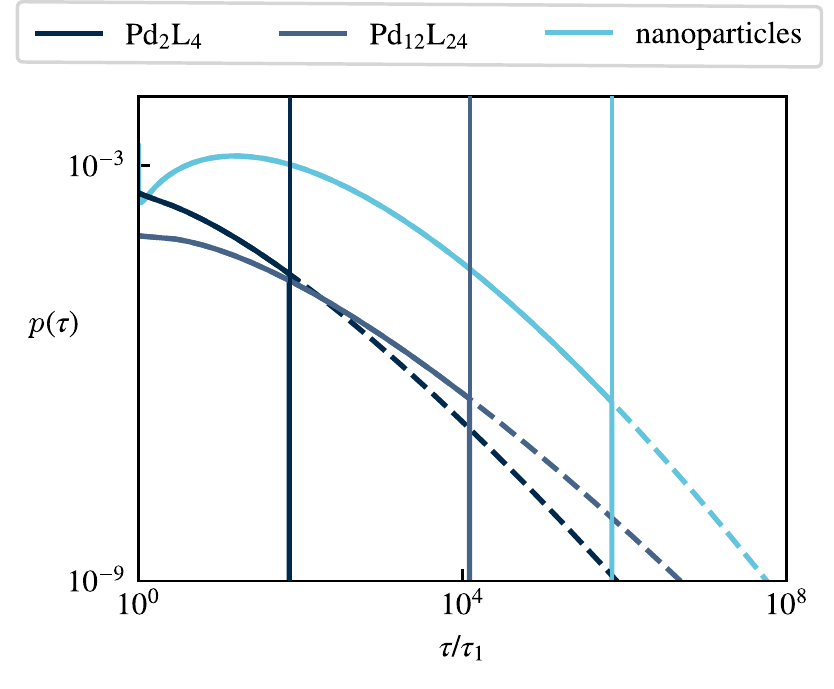}
    \caption{Distribution of relaxation times corresponding to the theoretical plots of Fig.~\ref{fig2}. The distribution associated with the Fe$^{3+}$ gel is a delta function for $\tau/\tau_1 = 1$ and is thus not represented on this graph. The time distributions are given by a Poisson distribution cut for $\tau = \tau_{N_\text{sat}}$ (vertical lines), as described in Eq.~\eqref{eq:response}. The dotted lines represent what these distributions would be in the absence of this saturation.}
    \label{fig:time_distribution}
\end{figure}

To further visualize the differences between the responses of our gels, we plot the distributions of relaxation times $p(\tau)$ for our fitted model in Fig.~\ref{fig:time_distribution}. The Fe$^{3+}$ gels, which relax according to a single exponential and whose $p(\tau)$ are therefore delta functions, are not represented there. In the Pd$_2$L$_4$ and Pd$_{12}$L$_{24}$ systems, a distribution characterized by an initially decreasing distribution of time scales is interrupted by a valence-dependent maximum relaxation time $\tau_{N_\text{sat}}$. That time is comprised within the range of time scales observed in Fig.~\ref{fig2}, accounting for the crossover to an exponential relaxation within this range. In nanoparticle systems, by contrast, the crossover occurs much later and thus cannot be directly observed in experiments. In all cases, the precise form of the distribution of time scales used in the domain $\tau<\tau_{N_\text{sat}}$ does not critically affect the predicted relaxation curves. Indeed, we show in Sec.~S6 of the Supplementary Information that replacing the Poisson distribution of Eq.~\eqref{eq10} with other distributions with the same mean and variance lead to essentially indistinguishable predictions over experimentally observable time scales. This emphasizes the robustness of our predictions to the details of that choice of distribution. They are instead primarily determined by the mean and maximum superbond sizes, $\bar{N}$ and $N_\text{sat}$.

In the limit of large $\bar{N}$ and even larger $N_\text{sat}$, the complex relaxation phase of our model characterized by $t<\tau_{N_\text{sat}}$ may display analytical behaviors identical to some widely used rheological fitting functions. In this regime, the Poisson distribution $p(N)$ of Eq.~\eqref{eq10} goes to a Gaussian. Since according to Eq.~\eqref{eq:tau_N} the variable $N$ is essentially the logarithm of the relaxation time $\tau$ for $\bar{N}\gg 1$, this results in a log-normal distribution of relaxation time scales:
\begin{align}\label{eq:log_normal}
    p(\tau)=&\frac{1}{\sqrt{2\pi\bar{N}}|\ln{p_\text{off}|}\tau}\\
    &\times\exp\left\lbrace -\frac{
    \left[\ln\tau+\ln{\frac{\bar{N}}{\tau_1}}+\ln p_\text{off}\left(\bar{N}-1\right)\right]^2
    }{2\bar{N}(\ln{p_\text{off}})^2}\right\rbrace.
    \nonumber
\end{align}
This result adds additional insights to this widely used fitting functional form, as it allows to relate the mean and variance of the distribution to the underlying crosslinker-scale parameters~\cite{epstein2019modulating,masurel2015role}. It moreover offers a potential molecular-level justification for its use in describing the complex relaxation of systems with multivalent crosslinks.
In the alternative case where $p(N)$ is a decaying exponential, our model results in power-law distributed relaxation time scales and the stress response function takes the form
\begin{equation}\label{eq:power_law}
    \sigma(t) \propto t^{-\gamma},
    \quad\text{with}\quad
    \gamma=\frac{1}{\bar{N}|\ln p_\text{off}|}.
\end{equation}
This result may also be presented in terms of the dependence of the storage and loss moduli on the frequency $\omega$ in an oscillatory rheology experiment. We thus predict that for $\gamma<1$
\begin{subequations}\label{eq:G_omega}
\begin{align}
    G'(\omega)&\approx
    \begin{cases}
        \omega^2 & \text{for }\omega\ll\tau_{N_\text{sat}}^{-1}\\
        \omega^{\gamma} & \text{for }\tau_{N_\text{sat}}^{-1}\ll\omega\ll\tau_1^{-1}\\
        \omega^{0} & \text{for }\tau_1^{-1}\ll\omega
    \end{cases}\\
        G''(\omega)&\approx
    \begin{cases}
        \omega^1 & \text{for }\omega\ll\tau_{N_\text{sat}}^{-1}\\
        \omega^{\gamma} & \text{for }\tau_{N_\text{sat}}^{-1}\ll\omega\ll\tau_1^{-1}\\
        \omega^{-1} & \text{for }\tau_1^{-1}\ll\omega
    \end{cases}
\end{align}
\end{subequations}
The results larger values of $\gamma$ and detailed derivations of Eqs.~(\ref{eq:log_normal}-\ref{eq:G_omega}) are shown in the Supplementary Information.
Again, this result has the potential to account for the power-law relaxation observed in many rheological systems~\cite{scott1959creep, keshavarz2017nonlinear, ballandPowerLawsMicrorheology2006}, in addition to providing a link to their microscopic constituents. Overall, these results suggest a possible control of the system's rheology through the characteristics of $p(N)$, which could in turn be modulated through the spatial distribution of the polymer strands and the dispersity of the crosslinkers.

\section*{Discussion}
Our simple model recapitulates a wide range of rheological behaviors in multivalent systems based on two key superbond parameters: the mean size $\bar{N}$ and the  maximum size $N_\text{sat}$. These respectively control the amplitude of the fluctuations in superbond size and the longest superbond relaxation time scale. Prior to the longest relaxation time, the system displays an increasingly nonexponential response for increasing $\bar{N}$ [Fig.~\ref{fig:stretched}(b)]. Beyond it, it crosses over into exponential relaxation. In contrast with widely used phenomenological fitting parameters, our two variables yield reliable insights into the underlying microscopic dynamics, as demonstrated by the agreement of their fitting values with our \textit{a priori} knownledge of four experimental systems covering a wide range of values of $\bar{N}$ and $N_\text{sat}$.

Our model bears a mathematical similarity with standard random energy trap models~\cite{bouchaud1992weak}. There, a long-tailed relaxation emerges from a short-tailed distribution of trap depths due to the exponential dependence of the relaxation times on the trap depths. Similarly, here a nonexponential relaxation emerges from a short-tailed distribution of superbond sizes $N$ [Eq.~\eqref{eq10}] thanks to the exponential dependence of $\tau_N$ on $N$ [Eq.~\eqref{eq:tau_N}]. In contrast with trap models, however, our model does not predict a glass transition upon a lowering of temperature. It instead displays a simple Arrhenius time-temperature relation, consistent with the experimental collapses in the insets of Fig.~\eqref{fig2}. Here again, an additional benefit of our approach is the direct connection between the predicted relaxation and experimentally accessible parameters such as the crosslinker surface (through $N_\text{sat}$).

Our model's focus on the collective aspects of superbond breaking and the characteristics of the crosslinkers implies that it encloses most of the physics of the polymer strands within a few mesoscopic parameters, mainly $\tau_0$ and $\Delta S$. Within our approach, the morphology of the polymer thus does not affect the form of our relaxation, although it may lead to a rescaling of the relaxation times of Eq.~\eqref{eq:tau_N}. This formulation remains valid as long as the length and concentration of the polymer strands is low enough that the polymer strands do not become significantly entangled, which could spoil the Poissonian attachment/detachment process of Eq.~\eqref{eq2}. Even in this case however this equation may not be strictly valid, as the polymer layer in a superbond with many bound crosslinkers tends to be more compressed than in one with few. This effect should leads to a smooth (likely power law) dependence of $\omega^\pm$ on $N$, which would preserve the dominance of the much more abrupt exponential dependence of $\tau_N$ on $N$. As a result, while such polymer brush effects could induce corrections in our estimations of the model parameters, the basic mechanism outlined here should still hold in their presence. 

Our model reproduces several qualitative characteristics of the rheology of multivalent gels, such as the strong influence of the crosslinker valence, Arrhenius temperature dependence and the transition between a nonexponential and an exponential regime at long times. Due to its simple, widely applicable microscopic assumptions, we believe that it could help shed light on and assist the design of a wide range of multivalent systems. Beyond composite gels, it could thus apply to RNA-protein biocondensates where multivalent interactions between proteins are mediated by RNA strands \cite{choiPhysicalPrinciplesUnderlying2020}, as well as cytoskeletal systems where filaments linked to many other filaments display a slow relaxation reminiscent of that of our multivalent crosslinkers~\cite{lielegSlowDynamicsInternal2011}.

\section*{Methods}
\subsection*{Materials}
4-arm poly(ethylene glycol) bis(acetic acid N-succinimidyl) ester (4-arm PEG-NHS) (MW = 10 kDa), 1-arm poly(ethylene glycol) bis(acetic acid N-succinimidyl) ester (1-arm PEG-NHS) (MW = 2000 Da) were purchased from JenKem Technology. Sodium sulfate ($\text{Na}_2\text{SO}_4$), sodium nitrite ($\text{NaNO}_2$), iron(III) acetylacetonate $(\text{Fe(acac)}_3)$, hydrochloric acid (HCl), dopamine hydrochloride, triethylamine (TEA), N-methylmorpholine (NMM), dimethyl sulfoxide (DMSO), methanol (MeOH), ethanol (EtOH), dichloromethane (DCM), N, N-dimethylformamide (DMF), diethyl ether ($\text{Et}_2\text{O}$), and chloroform ($\text{CHCl}_3$), were purchased from Sigma-Aldrich. All chemicals were used without further purification.

\subsection*{Synthesis of 1-arm PEG-Catechol (C)}
228 mg dopamine hydrochloride is neutralized for 15 min with 0.3 mL NMM in 7.5 mL dry DMF under $\text{N}_2$ atmosphere. Then 1 g mPEG-NHS (MW = 2000 Da) dissolved in 7.5 mL DMF is added, and the mixture is stirred with N$_2$ protection at room temperature for 24 hours. The reacted solution is acidified by adding 15 mL 1M HCl (aq), and the product is extracted with $\text{CHCl}_3$ 3 times. The organic layers are pooled together, dried with $\text{NaSO}_4$, and solvent is removed by rotary evaporation. Finally, the product concentrate is precipitated in cold $\text{Et}_2\text{O}$ (-20 $^\circ$C), filtered and dried. 1H NMR (300 MHz, D2O) $\delta$ (ppm): 6.7-6.8 (m, 3H, aromatic), 3.3-4.0 (m, $\text{-O-CH}_2\text{-CH}_2\text{-}$), 3.4 (t, 2H, CH$_2$ adjacent to aromatic ring), 2.7 (t, 2H, -CH$_2$-NH-CO-).

\subsection*{Synthesis of 4-arm PEG-Nitrocatechol (NC)}
178 mg of nitrodopamine hydrogen sulfate is neutralized for 15 min with 110 $\mu$L NMM in 4 mL dry DMF under N$_2$ atmosphere. Then 1 g 4-arm PEG-NHS (MW = 10 kDa) dissolved in 4 mL DMF is added, and the mixture is stirred with $\text{N}_2$ protection at room temperature for 24 hours. The reacted mixture is mixed with 15 mL 1M HCl$_{(aq)}$, dialyzed with water (MWCO = 3500 Da) for 2 days (water exchanged for more than 5 times), and freeze-dried. 1H NMR (300 MHz, D$_2$O) $\delta$ (ppm): 7.6 (m, 1H, aromatic), 6.7 (m, 1H, aromatic), 3.6-3.9 (m, -O-CH$_2$-CH$_2$-), 3.5 (t, 2H, CH$_2$ adjacent to aromatic ring), 3.1 (t, 2H, -CH$_2$-NH-CO-).

\subsection*{Synthesis of Fe$_3$O$_4$ nanoparticles (NPs)}
Bare Fe$_3$O$_4$ NPs are synthesized following previously reported methods (Li et al. 2016). 100 mg as-synthesized NPs are re-dispersed in 80 mL of 1:1 (v/v) solution of CHCl$_3$ and DMF, and 100 mg 1-arm PEG-C is added. The mixture is homogenized and equilibrated by pulsed sonication (pulse: 10 s on + 4 s off; power: 125 W) for 1 hour. Then the mixture is centrifuged at 10000 rpm for 10 min to remove any aggregates, and rotary evaporated at 50~$^\circ$C, 30 mbar to remove CHCl$_3$. Then the NP solution is precipitated in 150 mL cold Et$_2$O (-20 $^\circ$C). The precipitate is re-dispersed in H$_2$O, and freeze-dried. The resulting NPs are 7 nm in diameter.

\subsection*{Preparation of the Fe$3^+$-NC gels}
Preparation procedure is similar to a previously reported protocol~\cite{Holten-Andersen_PNAS_2010}, except that the gel is made in DMSO instead of H$_2$O. $50\,\mu$L of 200\,mg/mL 4-arm PEG-NC solution in DMSO is mixed with $16.7\,\mu$L of 80 mM FeCl$_3$ solution in DMSO (ligand: Fe$^{3+}$ molar ratio of 3:1). Then $33.3\,\mu$L DMSO and $13.8\,\mu$L TEA is added to facilitate deprotonation, and a gel is formed.

\subsection*{Preparation of the Pd$_2$L$_4$ gels}
The synthesis of polymer and gel preparation procedures for P2L4 is the same as a reported protocol~\cite{PolyMOC} with minor modifications. The annealing of the Pd$_2$L$_4$ polyMOC gel was done at 60°C for 1 hr instead of 80°C for 4 hr, and 1.05 equivalent of Pd(NO$_3$)$_2$. 2 H$_2$O (relative to bifunctional polymer ligand) was used instead of 1 equivalent.

\subsection*{Preparation of the Pd$_{12}$L$_{24}$ gels}
The synthesis of polymer and gel preparation procedures for polyMOC is the same as a reported protocol~\cite{PolyMOC}

\subsection*{Preparation of the NP gels}
Preparation procedure is the same as the reported protocol~\cite{songProgrammableAnisotropyPercolation2020}. Briefly, $20\,\text{mg}$ PEGylated Fe$_3$O$_4$ NPs (equivalent to $20\,\text{mg}$ Fe$_3$O$_4$ core) and $20\,\text{mg}$ 4-arm PEG-NC are mixed in a $0.2\,M$ HCl aqueous solution.  The solution mixture ($\text{pH} = 2$) is transferred into a mold and sealed, and a solid gel is obtained after curing in a 50 $^\circ$C oven for 24 hours.

\subsection*{Rheology}
Stress relaxation measurements are done on an Anton Paar rheometer with parallel plate geometry ($10\,\text{mm}$ diameter flat probe for NP gels and polyMOC gels, and $25\,\text{mm}$ diameter cone probe for Fe$^{3+}$ gels). All tests are done immediately after transferring the gel sample onto the sample stage. A Peltier hood is used for all experiments to control the measurement temperature and prevent solvent evaporation. H$_2$O based samples are furthermore sealed with mineral oil before experimentation to reduce the evaporation rate. Relaxation tests were performed by applying a $\gamma = 0.005$. step strain for the NP gel, and $\gamma = 0.02$ step strain for the other three systems.

\begin{acknowledgments}
HLR and ML thank Thibaut Divoux for valuable comments on the manuscript. This work was supported by Marie Curie Integration Grant PCIG12-GA-2012-334053, “Investissements d’Avenir” LabEx PALM (ANR-10-LABX-0039-PALM), ANR grants ANR-15-CE13-0004-03, ANR-21-CE11-0004-02, ANR-22-ERCC-0004-01 and ANR-22-CE30-0024-01, as well as ERC Starting Grant 677532 to ML. ML’s group belongs to the CNRS consortium AQV.
\end{acknowledgments}
\section*{Author contributions}
Hugo Le Roy and Martin Lenz designed the theoretical model and performed the theoretical analysis. Hugo Le Roy analyzed the data. Jake Song synthetized the ion and nanoparticle gels and carried out their rheological characterization under the supervision of Niels Holten-Andersen and Gareth H. McKinley.
Aleksandr Zhukhovitskiy and David Lunderberg synthetized the nanocage gels and carried out their rheological characterization under the supervision of Jeremiah A. Johnson.
The article was written by Hugo Le Roy, Jake Song and Martin Lenz. All others read and edited the manuscript.
\bibliographystyle{apsrev4-1}
\bibliography{valency_controls_non_exponential_relax}

\begin{thebibliography}{40}%
\makeatletter
\providecommand \@ifxundefined [1]{%
 \@ifx{#1\undefined}
}%
\providecommand \@ifnum [1]{%
 \ifnum #1\expandafter \@firstoftwo
 \else \expandafter \@secondoftwo
 \fi
}%
\providecommand \@ifx [1]{%
 \ifx #1\expandafter \@firstoftwo
 \else \expandafter \@secondoftwo
 \fi
}%
\providecommand \natexlab [1]{#1}%
\providecommand \enquote  [1]{``#1''}%
\providecommand \bibnamefont  [1]{#1}%
\providecommand \bibfnamefont [1]{#1}%
\providecommand \citenamefont [1]{#1}%
\providecommand \href@noop [0]{\@secondoftwo}%
\providecommand \href [0]{\begingroup \@sanitize@url \@href}%
\providecommand \@href[1]{\@@startlink{#1}\@@href}%
\providecommand \@@href[1]{\endgroup#1\@@endlink}%
\providecommand \@sanitize@url [0]{\catcode `\\12\catcode `\$12\catcode
  `\&12\catcode `\#12\catcode `\^12\catcode `\_12\catcode `\%12\relax}%
\providecommand \@@startlink[1]{}%
\providecommand \@@endlink[0]{}%
\providecommand \url  [0]{\begingroup\@sanitize@url \@url }%
\providecommand \@url [1]{\endgroup\@href {#1}{\urlprefix }}%
\providecommand \urlprefix  [0]{URL }%
\providecommand \Eprint [0]{\href }%
\providecommand \doibase [0]{http://dx.doi.org/}%
\providecommand \selectlanguage [0]{\@gobble}%
\providecommand \bibinfo  [0]{\@secondoftwo}%
\providecommand \bibfield  [0]{\@secondoftwo}%
\providecommand \translation [1]{[#1]}%
\providecommand \BibitemOpen [0]{}%
\providecommand \bibitemStop [0]{}%
\providecommand \bibitemNoStop [0]{.\EOS\space}%
\providecommand \EOS [0]{\spacefactor3000\relax}%
\providecommand \BibitemShut  [1]{\csname bibitem#1\endcsname}%
\let\auto@bib@innerbib\@empty
\bibitem [{\citenamefont {Ronsin}\ \emph {et~al.}(2016)\citenamefont {Ronsin},
  \citenamefont {Caroli},\ and\ \citenamefont {Baumberger}}]{Ronsin_2016}%
  \BibitemOpen
  \bibfield  {author} {\bibinfo {author} {\bibfnamefont {O.}~\bibnamefont
  {Ronsin}}, \bibinfo {author} {\bibfnamefont {C.}~\bibnamefont {Caroli}}, \
  and\ \bibinfo {author} {\bibfnamefont {T.}~\bibnamefont {Baumberger}},\
  }\href {\doibase 10.1063/1.4941456} {\bibfield  {journal} {\bibinfo
  {journal} {The Journal of Chemical Physics}\ }\textbf {\bibinfo {volume}
  {144}},\ \bibinfo {pages} {064904} (\bibinfo {year} {2016})}\BibitemShut
  {NoStop}%
\bibitem [{\citenamefont {Borges}\ \emph {et~al.}(2011)\citenamefont {Borges},
  \citenamefont {Eyholzer}, \citenamefont {Duc}, \citenamefont {Bourban},
  \citenamefont {Tingaut}, \citenamefont {Zimmermann}, \citenamefont
  {Pioletti},\ and\ \citenamefont {M{\aa}nson}}]{Ana_Acta_2011}%
  \BibitemOpen
  \bibfield  {author} {\bibinfo {author} {\bibfnamefont {A.~C.}\ \bibnamefont
  {Borges}}, \bibinfo {author} {\bibfnamefont {C.}~\bibnamefont {Eyholzer}},
  \bibinfo {author} {\bibfnamefont {F.}~\bibnamefont {Duc}}, \bibinfo {author}
  {\bibfnamefont {P.-E.}\ \bibnamefont {Bourban}}, \bibinfo {author}
  {\bibfnamefont {P.}~\bibnamefont {Tingaut}}, \bibinfo {author} {\bibfnamefont
  {T.}~\bibnamefont {Zimmermann}}, \bibinfo {author} {\bibfnamefont {D.~P.}\
  \bibnamefont {Pioletti}}, \ and\ \bibinfo {author} {\bibfnamefont {J.-A.~E.}\
  \bibnamefont {M{\aa}nson}},\ }\href {\doibase 10.1016/j.actbio.2011.05.029}
  {\bibfield  {journal} {\bibinfo  {journal} {Acta Biomaterialia}\ }\textbf
  {\bibinfo {volume} {7}},\ \bibinfo {pages} {3412} (\bibinfo {year}
  {2011})}\BibitemShut {NoStop}%
\bibitem [{\citenamefont {Chaudhuri}\ \emph {et~al.}(2015)\citenamefont
  {Chaudhuri}, \citenamefont {Gu}, \citenamefont {Darnell}, \citenamefont
  {Klumpers}, \citenamefont {Bencherif}, \citenamefont {Weaver}, \citenamefont
  {Huebsch},\ and\ \citenamefont {Mooney}}]{chaudhuri2015substrate}%
  \BibitemOpen
  \bibfield  {author} {\bibinfo {author} {\bibfnamefont {O.}~\bibnamefont
  {Chaudhuri}}, \bibinfo {author} {\bibfnamefont {L.}~\bibnamefont {Gu}},
  \bibinfo {author} {\bibfnamefont {M.}~\bibnamefont {Darnell}}, \bibinfo
  {author} {\bibfnamefont {D.}~\bibnamefont {Klumpers}}, \bibinfo {author}
  {\bibfnamefont {S.~A.}\ \bibnamefont {Bencherif}}, \bibinfo {author}
  {\bibfnamefont {J.~C.}\ \bibnamefont {Weaver}}, \bibinfo {author}
  {\bibfnamefont {N.}~\bibnamefont {Huebsch}}, \ and\ \bibinfo {author}
  {\bibfnamefont {D.~J.}\ \bibnamefont {Mooney}},\ }\href {\doibase
  10.1038/ncomms7365} {\bibfield  {journal} {\bibinfo  {journal} {Nature
  Communications}\ }\textbf {\bibinfo {volume} {6}},\ \bibinfo {pages} {6365}
  (\bibinfo {year} {2015})}\BibitemShut {NoStop}%
\bibitem [{\citenamefont {Kollmannsberger}\ and\ \citenamefont
  {Fabry}(2011)}]{kollmannsberger2011linear}%
  \BibitemOpen
  \bibfield  {author} {\bibinfo {author} {\bibfnamefont {P.}~\bibnamefont
  {Kollmannsberger}}\ and\ \bibinfo {author} {\bibfnamefont {B.}~\bibnamefont
  {Fabry}},\ }\href@noop {} {\bibfield  {journal} {\bibinfo  {journal} {Annual
  review of materials research}\ }\textbf {\bibinfo {volume} {41}},\ \bibinfo
  {pages} {75} (\bibinfo {year} {2011})}\BibitemShut {NoStop}%
\bibitem [{\citenamefont {Chaudhuri}\ \emph {et~al.}(2020)\citenamefont
  {Chaudhuri}, \citenamefont {Cooper-White}, \citenamefont {Janmey},
  \citenamefont {Mooney},\ and\ \citenamefont {Shenoy}}]{chaudhuri2020effects}%
  \BibitemOpen
  \bibfield  {author} {\bibinfo {author} {\bibfnamefont {O.}~\bibnamefont
  {Chaudhuri}}, \bibinfo {author} {\bibfnamefont {J.}~\bibnamefont
  {Cooper-White}}, \bibinfo {author} {\bibfnamefont {P.~A.}\ \bibnamefont
  {Janmey}}, \bibinfo {author} {\bibfnamefont {D.~J.}\ \bibnamefont {Mooney}},
  \ and\ \bibinfo {author} {\bibfnamefont {V.~B.}\ \bibnamefont {Shenoy}},\
  }\href@noop {} {\bibfield  {journal} {\bibinfo  {journal} {Nature}\ }\textbf
  {\bibinfo {volume} {584}},\ \bibinfo {pages} {535} (\bibinfo {year}
  {2020})}\BibitemShut {NoStop}%
\bibitem [{\citenamefont {Celli}\ \emph {et~al.}(2009)\citenamefont {Celli},
  \citenamefont {Turner}, \citenamefont {Afdhal}, \citenamefont {Keates},
  \citenamefont {Ghiran}, \citenamefont {Kelly}, \citenamefont {Ewoldt},
  \citenamefont {McKinley}, \citenamefont {So}, \citenamefont {Erramilli} \emph
  {et~al.}}]{celli2009helicobacter}%
  \BibitemOpen
  \bibfield  {author} {\bibinfo {author} {\bibfnamefont {J.~P.}\ \bibnamefont
  {Celli}}, \bibinfo {author} {\bibfnamefont {B.~S.}\ \bibnamefont {Turner}},
  \bibinfo {author} {\bibfnamefont {N.~H.}\ \bibnamefont {Afdhal}}, \bibinfo
  {author} {\bibfnamefont {S.}~\bibnamefont {Keates}}, \bibinfo {author}
  {\bibfnamefont {I.}~\bibnamefont {Ghiran}}, \bibinfo {author} {\bibfnamefont
  {C.~P.}\ \bibnamefont {Kelly}}, \bibinfo {author} {\bibfnamefont {R.~H.}\
  \bibnamefont {Ewoldt}}, \bibinfo {author} {\bibfnamefont {G.~H.}\
  \bibnamefont {McKinley}}, \bibinfo {author} {\bibfnamefont {P.}~\bibnamefont
  {So}}, \bibinfo {author} {\bibfnamefont {S.}~\bibnamefont {Erramilli}},
  \emph {et~al.},\ }\href@noop {} {\bibfield  {journal} {\bibinfo  {journal}
  {Proceedings of the National Academy of Sciences}\ }\textbf {\bibinfo
  {volume} {106}},\ \bibinfo {pages} {14321} (\bibinfo {year}
  {2009})}\BibitemShut {NoStop}%
\bibitem [{\citenamefont {Fabbri}\ and\ \citenamefont
  {Stoodley}(2016)}]{fabbri2016mechanical}%
  \BibitemOpen
  \bibfield  {author} {\bibinfo {author} {\bibfnamefont {S.}~\bibnamefont
  {Fabbri}}\ and\ \bibinfo {author} {\bibfnamefont {P.}~\bibnamefont
  {Stoodley}},\ }\href@noop {} {\bibfield  {journal} {\bibinfo  {journal} {The
  perfect slime—microbial extracellular polymeric substances}\ ,\ \bibinfo
  {pages} {153}} (\bibinfo {year} {2016})}\BibitemShut {NoStop}%
\bibitem [{\citenamefont {Song}\ \emph {et~al.}(2022)\citenamefont {Song},
  \citenamefont {Holten-Andersen},\ and\ \citenamefont
  {McKinley}}]{song2022non}%
  \BibitemOpen
  \bibfield  {author} {\bibinfo {author} {\bibfnamefont {J.}~\bibnamefont
  {Song}}, \bibinfo {author} {\bibfnamefont {N.}~\bibnamefont
  {Holten-Andersen}}, \ and\ \bibinfo {author} {\bibfnamefont {G.~H.}\
  \bibnamefont {McKinley}},\ }\href@noop {} {\bibfield  {journal} {\bibinfo
  {journal} {arXiv preprint arXiv:2211.06438}\ } (\bibinfo {year}
  {2022})}\BibitemShut {NoStop}%
\bibitem [{\citenamefont {Bouchaud}(2008)}]{Bouchaud_stretch_exponential}%
  \BibitemOpen
  \bibfield  {author} {\bibinfo {author} {\bibfnamefont {J.-P.}\ \bibnamefont
  {Bouchaud}},\ }in\ \href {\doibase 10.1002/9783527622979.ch11} {\emph
  {\bibinfo {booktitle} {Anomalous {{Transport}}}}}\ (\bibinfo  {publisher}
  {{John Wiley \& Sons, Ltd}},\ \bibinfo {year} {2008})\ Chap.~\bibinfo
  {chapter} {11}, pp.\ \bibinfo {pages} {327--345}\BibitemShut {NoStop}%
\bibitem [{\citenamefont {Blair}\ and\ \citenamefont
  {Burnett}(1959)}]{scott1959creep}%
  \BibitemOpen
  \bibfield  {author} {\bibinfo {author} {\bibfnamefont {G.~S.}\ \bibnamefont
  {Blair}}\ and\ \bibinfo {author} {\bibfnamefont {J.}~\bibnamefont
  {Burnett}},\ }\href {\doibase 10.1088/0508-3443/10/1/304} {\bibfield
  {journal} {\bibinfo  {journal} {British Journal of Applied Physics}\ }\textbf
  {\bibinfo {volume} {10}},\ \bibinfo {pages} {15} (\bibinfo {year}
  {1959})}\BibitemShut {NoStop}%
\bibitem [{\citenamefont {Keshavarz}\ \emph {et~al.}(2017)\citenamefont
  {Keshavarz}, \citenamefont {Divoux}, \citenamefont {Manneville},\ and\
  \citenamefont {McKinley}}]{keshavarz2017nonlinear}%
  \BibitemOpen
  \bibfield  {author} {\bibinfo {author} {\bibfnamefont {B.}~\bibnamefont
  {Keshavarz}}, \bibinfo {author} {\bibfnamefont {T.}~\bibnamefont {Divoux}},
  \bibinfo {author} {\bibfnamefont {S.}~\bibnamefont {Manneville}}, \ and\
  \bibinfo {author} {\bibfnamefont {G.~H.}\ \bibnamefont {McKinley}},\ }\href
  {\doibase 10.1021/acsmacrolett.7b00213} {\bibfield  {journal} {\bibinfo
  {journal} {ACS Macro Letters}\ }\textbf {\bibinfo {volume} {6}},\ \bibinfo
  {pages} {663} (\bibinfo {year} {2017})}\BibitemShut {NoStop}%
\bibitem [{\citenamefont {Balland}\ \emph {et~al.}(2006)\citenamefont
  {Balland}, \citenamefont {Desprat}, \citenamefont {Icard}, \citenamefont
  {F{\'e}r{\'e}ol}, \citenamefont {Asnacios}, \citenamefont {Browaeys},
  \citenamefont {H{\'e}non},\ and\ \citenamefont
  {Gallet}}]{ballandPowerLawsMicrorheology2006}%
  \BibitemOpen
  \bibfield  {author} {\bibinfo {author} {\bibfnamefont {M.}~\bibnamefont
  {Balland}}, \bibinfo {author} {\bibfnamefont {N.}~\bibnamefont {Desprat}},
  \bibinfo {author} {\bibfnamefont {D.}~\bibnamefont {Icard}}, \bibinfo
  {author} {\bibfnamefont {S.}~\bibnamefont {F{\'e}r{\'e}ol}}, \bibinfo
  {author} {\bibfnamefont {A.}~\bibnamefont {Asnacios}}, \bibinfo {author}
  {\bibfnamefont {J.}~\bibnamefont {Browaeys}}, \bibinfo {author}
  {\bibfnamefont {S.}~\bibnamefont {H{\'e}non}}, \ and\ \bibinfo {author}
  {\bibfnamefont {F.}~\bibnamefont {Gallet}},\ }\href {\doibase
  10.1103/PhysRevE.74.021911} {\bibfield  {journal} {\bibinfo  {journal}
  {Physical Review E}\ }\textbf {\bibinfo {volume} {74}},\ \bibinfo {pages}
  {021911} (\bibinfo {year} {2006})}\BibitemShut {NoStop}%
\bibitem [{\citenamefont {Epstein}\ \emph {et~al.}(2019)\citenamefont
  {Epstein}, \citenamefont {Martinetti}, \citenamefont {Kollarigowda},
  \citenamefont {Carey-De La~Torre}, \citenamefont {Moore}, \citenamefont
  {Ewoldt},\ and\ \citenamefont {Braun}}]{epstein2019modulating}%
  \BibitemOpen
  \bibfield  {author} {\bibinfo {author} {\bibfnamefont {E.~S.}\ \bibnamefont
  {Epstein}}, \bibinfo {author} {\bibfnamefont {L.}~\bibnamefont {Martinetti}},
  \bibinfo {author} {\bibfnamefont {R.~H.}\ \bibnamefont {Kollarigowda}},
  \bibinfo {author} {\bibfnamefont {O.}~\bibnamefont {Carey-De La~Torre}},
  \bibinfo {author} {\bibfnamefont {J.~S.}\ \bibnamefont {Moore}}, \bibinfo
  {author} {\bibfnamefont {R.~H.}\ \bibnamefont {Ewoldt}}, \ and\ \bibinfo
  {author} {\bibfnamefont {P.~V.}\ \bibnamefont {Braun}},\ }\href@noop {}
  {\bibfield  {journal} {\bibinfo  {journal} {Journal of the American Chemical
  Society}\ }\textbf {\bibinfo {volume} {141}},\ \bibinfo {pages} {3597}
  (\bibinfo {year} {2019})}\BibitemShut {NoStop}%
\bibitem [{\citenamefont {Masurel}\ \emph {et~al.}(2015)\citenamefont
  {Masurel}, \citenamefont {Cantournet}, \citenamefont {Dequidt}, \citenamefont
  {Long}, \citenamefont {Montes},\ and\ \citenamefont
  {Lequeux}}]{masurel2015role}%
  \BibitemOpen
  \bibfield  {author} {\bibinfo {author} {\bibfnamefont {R.~J.}\ \bibnamefont
  {Masurel}}, \bibinfo {author} {\bibfnamefont {S.}~\bibnamefont {Cantournet}},
  \bibinfo {author} {\bibfnamefont {A.}~\bibnamefont {Dequidt}}, \bibinfo
  {author} {\bibfnamefont {D.~R.}\ \bibnamefont {Long}}, \bibinfo {author}
  {\bibfnamefont {H.}~\bibnamefont {Montes}}, \ and\ \bibinfo {author}
  {\bibfnamefont {F.}~\bibnamefont {Lequeux}},\ }\href@noop {} {\bibfield
  {journal} {\bibinfo  {journal} {Macromolecules}\ }\textbf {\bibinfo {volume}
  {48}},\ \bibinfo {pages} {6690} (\bibinfo {year} {2015})}\BibitemShut
  {NoStop}%
\bibitem [{\citenamefont {Alberto~Parada}\ and\ \citenamefont
  {Zhao}(2018)}]{parada2018ideal}%
  \BibitemOpen
  \bibfield  {author} {\bibinfo {author} {\bibfnamefont {G.}~\bibnamefont
  {Alberto~Parada}}\ and\ \bibinfo {author} {\bibfnamefont {X.}~\bibnamefont
  {Zhao}},\ }\href {\doibase 10.1039/C8SM00646F} {\bibfield  {journal}
  {\bibinfo  {journal} {Soft Matter}\ }\textbf {\bibinfo {volume} {14}},\
  \bibinfo {pages} {5186} (\bibinfo {year} {2018})}\BibitemShut {NoStop}%
\bibitem [{\citenamefont {Grindy}\ \emph {et~al.}(2015)\citenamefont {Grindy},
  \citenamefont {Learsch}, \citenamefont {Mozhdehi}, \citenamefont {Cheng},
  \citenamefont {Barrett}, \citenamefont {Guan}, \citenamefont {Messersmith},\
  and\ \citenamefont {{Holten-Andersen}}}]{Grindy_Nat_Mat_2016}%
  \BibitemOpen
  \bibfield  {author} {\bibinfo {author} {\bibfnamefont {S.~C.}\ \bibnamefont
  {Grindy}}, \bibinfo {author} {\bibfnamefont {R.}~\bibnamefont {Learsch}},
  \bibinfo {author} {\bibfnamefont {D.}~\bibnamefont {Mozhdehi}}, \bibinfo
  {author} {\bibfnamefont {J.}~\bibnamefont {Cheng}}, \bibinfo {author}
  {\bibfnamefont {D.~G.}\ \bibnamefont {Barrett}}, \bibinfo {author}
  {\bibfnamefont {Z.}~\bibnamefont {Guan}}, \bibinfo {author} {\bibfnamefont
  {P.~B.}\ \bibnamefont {Messersmith}}, \ and\ \bibinfo {author} {\bibfnamefont
  {N.}~\bibnamefont {{Holten-Andersen}}},\ }\href {\doibase 10.1038/nmat4401}
  {\bibfield  {journal} {\bibinfo  {journal} {Nature Materials}\ }\textbf
  {\bibinfo {volume} {14}},\ \bibinfo {pages} {1210} (\bibinfo {year}
  {2015})}\BibitemShut {NoStop}%
\bibitem [{\citenamefont {Rosales}\ and\ \citenamefont
  {Anseth}(2016)}]{rosales2016design}%
  \BibitemOpen
  \bibfield  {author} {\bibinfo {author} {\bibfnamefont {A.~M.}\ \bibnamefont
  {Rosales}}\ and\ \bibinfo {author} {\bibfnamefont {K.~S.}\ \bibnamefont
  {Anseth}},\ }\href@noop {} {\bibfield  {journal} {\bibinfo  {journal} {Nature
  Reviews Materials}\ }\textbf {\bibinfo {volume} {1}},\ \bibinfo {pages} {1}
  (\bibinfo {year} {2016})}\BibitemShut {NoStop}%
\bibitem [{\citenamefont {Semenov}\ \emph {et~al.}(1995)\citenamefont
  {Semenov}, \citenamefont {Joanny},\ and\ \citenamefont
  {Khokhlov}}]{semenovAssociatingPolymersEquilibrium1995}%
  \BibitemOpen
  \bibfield  {author} {\bibinfo {author} {\bibfnamefont {A.~N.}\ \bibnamefont
  {Semenov}}, \bibinfo {author} {\bibfnamefont {J.-F.}\ \bibnamefont {Joanny}},
  \ and\ \bibinfo {author} {\bibfnamefont {A.~R.}\ \bibnamefont {Khokhlov}},\
  }\href {\doibase 10.1021/ma00108a038} {\bibfield  {journal} {\bibinfo
  {journal} {Macromolecules}\ }\textbf {\bibinfo {volume} {28}},\ \bibinfo
  {pages} {1066} (\bibinfo {year} {1995})}\BibitemShut {NoStop}%
\bibitem [{\citenamefont {Michel}\ \emph {et~al.}(2001)\citenamefont {Michel},
  \citenamefont {Appell}, \citenamefont {Molino}, \citenamefont {Kieffer},\
  and\ \citenamefont {Porte}}]{michelUnstableFlowNonmonotonic2001}%
  \BibitemOpen
  \bibfield  {author} {\bibinfo {author} {\bibfnamefont {E.}~\bibnamefont
  {Michel}}, \bibinfo {author} {\bibfnamefont {J.}~\bibnamefont {Appell}},
  \bibinfo {author} {\bibfnamefont {F.}~\bibnamefont {Molino}}, \bibinfo
  {author} {\bibfnamefont {J.}~\bibnamefont {Kieffer}}, \ and\ \bibinfo
  {author} {\bibfnamefont {G.}~\bibnamefont {Porte}},\ }\href {\doibase
  10.1122/1.1413507} {\bibfield  {journal} {\bibinfo  {journal} {Journal of
  Rheology}\ }\textbf {\bibinfo {volume} {45}},\ \bibinfo {pages} {1465}
  (\bibinfo {year} {2001})}\BibitemShut {NoStop}%
\bibitem [{\citenamefont {Li}\ \emph {et~al.}(2016)\citenamefont {Li},
  \citenamefont {Barrett}, \citenamefont {Messersmith},\ and\ \citenamefont
  {Holten-Andersen}}]{li2016controlling}%
  \BibitemOpen
  \bibfield  {author} {\bibinfo {author} {\bibfnamefont {Q.}~\bibnamefont
  {Li}}, \bibinfo {author} {\bibfnamefont {D.~G.}\ \bibnamefont {Barrett}},
  \bibinfo {author} {\bibfnamefont {P.~B.}\ \bibnamefont {Messersmith}}, \ and\
  \bibinfo {author} {\bibfnamefont {N.}~\bibnamefont {Holten-Andersen}},\
  }\href@noop {} {\bibfield  {journal} {\bibinfo  {journal} {ACS nano}\
  }\textbf {\bibinfo {volume} {10}},\ \bibinfo {pages} {1317} (\bibinfo {year}
  {2016})}\BibitemShut {NoStop}%
\bibitem [{\citenamefont {Zhukhovitskiy}\ \emph
  {et~al.}(2016{\natexlab{a}})\citenamefont {Zhukhovitskiy}, \citenamefont
  {Zhao}, \citenamefont {Zhong}, \citenamefont {Keeler}, \citenamefont {Alt},
  \citenamefont {Teichen}, \citenamefont {Griffin}, \citenamefont {Hore},
  \citenamefont {Willard},\ and\ \citenamefont
  {Johnson}}]{zhukhovitskiyPolymerStructureDependent2016}%
  \BibitemOpen
  \bibfield  {author} {\bibinfo {author} {\bibfnamefont {A.~V.}\ \bibnamefont
  {Zhukhovitskiy}}, \bibinfo {author} {\bibfnamefont {J.}~\bibnamefont {Zhao}},
  \bibinfo {author} {\bibfnamefont {M.}~\bibnamefont {Zhong}}, \bibinfo
  {author} {\bibfnamefont {E.~G.}\ \bibnamefont {Keeler}}, \bibinfo {author}
  {\bibfnamefont {E.~A.}\ \bibnamefont {Alt}}, \bibinfo {author} {\bibfnamefont
  {P.}~\bibnamefont {Teichen}}, \bibinfo {author} {\bibfnamefont {R.~G.}\
  \bibnamefont {Griffin}}, \bibinfo {author} {\bibfnamefont {M.~J.~A.}\
  \bibnamefont {Hore}}, \bibinfo {author} {\bibfnamefont {A.~P.}\ \bibnamefont
  {Willard}}, \ and\ \bibinfo {author} {\bibfnamefont {J.~A.}\ \bibnamefont
  {Johnson}},\ }\href {\doibase 10.1021/acs.macromol.6b01607} {\bibfield
  {journal} {\bibinfo  {journal} {Macromolecules}\ }\textbf {\bibinfo {volume}
  {49}},\ \bibinfo {pages} {6896} (\bibinfo {year}
  {2016}{\natexlab{a}})}\BibitemShut {NoStop}%
\bibitem [{\citenamefont {Wang}\ \emph {et~al.}(2010)\citenamefont {Wang},
  \citenamefont {Mynar}, \citenamefont {Yoshida}, \citenamefont {Lee},
  \citenamefont {Lee}, \citenamefont {Okuro}, \citenamefont {Kinbara},\ and\
  \citenamefont {Aida}}]{wang2010high}%
  \BibitemOpen
  \bibfield  {author} {\bibinfo {author} {\bibfnamefont {Q.}~\bibnamefont
  {Wang}}, \bibinfo {author} {\bibfnamefont {J.~L.}\ \bibnamefont {Mynar}},
  \bibinfo {author} {\bibfnamefont {M.}~\bibnamefont {Yoshida}}, \bibinfo
  {author} {\bibfnamefont {E.}~\bibnamefont {Lee}}, \bibinfo {author}
  {\bibfnamefont {M.}~\bibnamefont {Lee}}, \bibinfo {author} {\bibfnamefont
  {K.}~\bibnamefont {Okuro}}, \bibinfo {author} {\bibfnamefont
  {K.}~\bibnamefont {Kinbara}}, \ and\ \bibinfo {author} {\bibfnamefont
  {T.}~\bibnamefont {Aida}},\ }\href {\doibase 10.1038/nature08693} {\bibfield
  {journal} {\bibinfo  {journal} {Nature}\ }\textbf {\bibinfo {volume} {463}},\
  \bibinfo {pages} {339} (\bibinfo {year} {2010})}\BibitemShut {NoStop}%
\bibitem [{\citenamefont {Chatterjee}\ \emph {et~al.}(2014)\citenamefont
  {Chatterjee}, \citenamefont {Nakatani},\ and\ \citenamefont
  {Van~Dyk}}]{Chatterjee_MacroMol_2014}%
  \BibitemOpen
  \bibfield  {author} {\bibinfo {author} {\bibfnamefont {T.}~\bibnamefont
  {Chatterjee}}, \bibinfo {author} {\bibfnamefont {A.~I.}\ \bibnamefont
  {Nakatani}}, \ and\ \bibinfo {author} {\bibfnamefont {A.~K.}\ \bibnamefont
  {Van~Dyk}},\ }\href {\doibase 10.1021/ma401566k} {\bibfield  {journal}
  {\bibinfo  {journal} {Macromolecules}\ }\textbf {\bibinfo {volume} {47}},\
  \bibinfo {pages} {1155} (\bibinfo {year} {2014})}\BibitemShut {NoStop}%
\bibitem [{\citenamefont {Flory}(1953)}]{flory1953principles}%
  \BibitemOpen
  \bibfield  {author} {\bibinfo {author} {\bibfnamefont {P.~J.}\ \bibnamefont
  {Flory}},\ }\href@noop {} {\emph {\bibinfo {title} {Principles of polymer
  chemistry}}}\ (\bibinfo  {publisher} {Cornell university press},\ \bibinfo
  {year} {1953})\BibitemShut {NoStop}%
\bibitem [{\citenamefont {{Gomez-Casado}}\ \emph {et~al.}(2011)\citenamefont
  {{Gomez-Casado}}, \citenamefont {Dam}, \citenamefont {Yilmaz}, \citenamefont
  {Florea}, \citenamefont {Jonkheijm},\ and\ \citenamefont
  {Huskens}}]{gomez2011probing}%
  \BibitemOpen
  \bibfield  {author} {\bibinfo {author} {\bibfnamefont {A.}~\bibnamefont
  {{Gomez-Casado}}}, \bibinfo {author} {\bibfnamefont {H.~H.}\ \bibnamefont
  {Dam}}, \bibinfo {author} {\bibfnamefont {M.~D.}\ \bibnamefont {Yilmaz}},
  \bibinfo {author} {\bibfnamefont {D.}~\bibnamefont {Florea}}, \bibinfo
  {author} {\bibfnamefont {P.}~\bibnamefont {Jonkheijm}}, \ and\ \bibinfo
  {author} {\bibfnamefont {J.}~\bibnamefont {Huskens}},\ }\href {\doibase
  10.1021/ja2016125} {\bibfield  {journal} {\bibinfo  {journal} {Journal of the
  American Chemical Society}\ }\textbf {\bibinfo {volume} {133}},\ \bibinfo
  {pages} {10849} (\bibinfo {year} {2011})}\BibitemShut {NoStop}%
\bibitem [{\citenamefont {Bertin}\ and\ \citenamefont
  {Bouchaud}(2003)}]{bertinSubdiffusionLocalizationOnedimensional2003}%
  \BibitemOpen
  \bibfield  {author} {\bibinfo {author} {\bibfnamefont {E.~M.}\ \bibnamefont
  {Bertin}}\ and\ \bibinfo {author} {\bibfnamefont {J.-P.}\ \bibnamefont
  {Bouchaud}},\ }\href {\doibase 10.1103/PhysRevE.67.026128} {\bibfield
  {journal} {\bibinfo  {journal} {Physical Review E}\ }\textbf {\bibinfo
  {volume} {67}},\ \bibinfo {pages} {026128} (\bibinfo {year}
  {2003})}\BibitemShut {NoStop}%
\bibitem [{\citenamefont {Sollich}\ \emph {et~al.}(1997)\citenamefont
  {Sollich}, \citenamefont {Lequeux}, \citenamefont {H{\'e}braud},\ and\
  \citenamefont {Cates}}]{SGR}%
  \BibitemOpen
  \bibfield  {author} {\bibinfo {author} {\bibfnamefont {P.}~\bibnamefont
  {Sollich}}, \bibinfo {author} {\bibfnamefont {F.}~\bibnamefont {Lequeux}},
  \bibinfo {author} {\bibfnamefont {P.}~\bibnamefont {H{\'e}braud}}, \ and\
  \bibinfo {author} {\bibfnamefont {M.~E.}\ \bibnamefont {Cates}},\ }\href
  {\doibase 10.1103/PhysRevLett.78.2020} {\bibfield  {journal} {\bibinfo
  {journal} {Physical Review Letters}\ }\textbf {\bibinfo {volume} {78}},\
  \bibinfo {pages} {2020} (\bibinfo {year} {1997})}\BibitemShut {NoStop}%
\bibitem [{\citenamefont {Trachenko}\ and\ \citenamefont
  {Zaccone}(2021)}]{trachenkoSlowStretchedexponentialFast2021}%
  \BibitemOpen
  \bibfield  {author} {\bibinfo {author} {\bibfnamefont {K.}~\bibnamefont
  {Trachenko}}\ and\ \bibinfo {author} {\bibfnamefont {A.}~\bibnamefont
  {Zaccone}},\ }\href {\doibase 10.1088/1361-648X/ac04cd} {\bibfield  {journal}
  {\bibinfo  {journal} {Journal of Physics: Condensed Matter}\ }\textbf
  {\bibinfo {volume} {33}},\ \bibinfo {pages} {315101} (\bibinfo {year}
  {2021})}\BibitemShut {NoStop}%
\bibitem [{\citenamefont {Raspaud}\ \emph {et~al.}(1996)\citenamefont
  {Raspaud}, \citenamefont {Lairez}, \citenamefont {Adam},\ and\ \citenamefont
  {Carton}}]{raspaudTriblockCopolymersSelective1996}%
  \BibitemOpen
  \bibfield  {author} {\bibinfo {author} {\bibfnamefont {E.}~\bibnamefont
  {Raspaud}}, \bibinfo {author} {\bibfnamefont {D.}~\bibnamefont {Lairez}},
  \bibinfo {author} {\bibfnamefont {M.}~\bibnamefont {Adam}}, \ and\ \bibinfo
  {author} {\bibfnamefont {J.-P.}\ \bibnamefont {Carton}},\ }\href {\doibase
  10.1021/ma951172x} {\bibfield  {journal} {\bibinfo  {journal}
  {Macromolecules}\ }\textbf {\bibinfo {volume} {29}},\ \bibinfo {pages} {1269}
  (\bibinfo {year} {1996})}\BibitemShut {NoStop}%
\bibitem [{\citenamefont {Texier}(2000)}]{texierIndividualEnergyLevel2000}%
  \BibitemOpen
  \bibfield  {author} {\bibinfo {author} {\bibfnamefont {C.}~\bibnamefont
  {Texier}},\ }\href {\doibase 10.1088/0305-4470/33/35/303} {\bibfield
  {journal} {\bibinfo  {journal} {Journal of Physics A: Mathematical and
  General}\ }\textbf {\bibinfo {volume} {33}},\ \bibinfo {pages} {6095}
  (\bibinfo {year} {2000})}\BibitemShut {NoStop}%
\bibitem [{\citenamefont {Zhukhovitskiy}\ \emph
  {et~al.}(2016{\natexlab{b}})\citenamefont {Zhukhovitskiy}, \citenamefont
  {Zhong}, \citenamefont {Keeler}, \citenamefont {Michaelis}, \citenamefont
  {Sun}, \citenamefont {Hore}, \citenamefont {Pochan}, \citenamefont {Griffin},
  \citenamefont {Willard},\ and\ \citenamefont {Johnson}}]{PolyMOC}%
  \BibitemOpen
  \bibfield  {author} {\bibinfo {author} {\bibfnamefont {A.~V.}\ \bibnamefont
  {Zhukhovitskiy}}, \bibinfo {author} {\bibfnamefont {M.}~\bibnamefont
  {Zhong}}, \bibinfo {author} {\bibfnamefont {E.~G.}\ \bibnamefont {Keeler}},
  \bibinfo {author} {\bibfnamefont {V.~K.}\ \bibnamefont {Michaelis}}, \bibinfo
  {author} {\bibfnamefont {J.~E.~P.}\ \bibnamefont {Sun}}, \bibinfo {author}
  {\bibfnamefont {M.~J.~A.}\ \bibnamefont {Hore}}, \bibinfo {author}
  {\bibfnamefont {D.~J.}\ \bibnamefont {Pochan}}, \bibinfo {author}
  {\bibfnamefont {R.~G.}\ \bibnamefont {Griffin}}, \bibinfo {author}
  {\bibfnamefont {A.~P.}\ \bibnamefont {Willard}}, \ and\ \bibinfo {author}
  {\bibfnamefont {J.~A.}\ \bibnamefont {Johnson}},\ }\href {\doibase
  10.1038/nchem.2390} {\bibfield  {journal} {\bibinfo  {journal} {Nature
  Chemistry}\ }\textbf {\bibinfo {volume} {8}},\ \bibinfo {pages} {33}
  (\bibinfo {year} {2016}{\natexlab{b}})}\BibitemShut {NoStop}%
\bibitem [{\citenamefont {Hsiao}\ \emph {et~al.}(2012)\citenamefont {Hsiao},
  \citenamefont {Newman}, \citenamefont {Glotzer},\ and\ \citenamefont
  {Solomon}}]{hsiaoRoleIsostaticityLoadbearing2012}%
  \BibitemOpen
  \bibfield  {author} {\bibinfo {author} {\bibfnamefont {L.~C.}\ \bibnamefont
  {Hsiao}}, \bibinfo {author} {\bibfnamefont {R.~S.}\ \bibnamefont {Newman}},
  \bibinfo {author} {\bibfnamefont {S.~C.}\ \bibnamefont {Glotzer}}, \ and\
  \bibinfo {author} {\bibfnamefont {M.~J.}\ \bibnamefont {Solomon}},\ }\href
  {\doibase 10.1073/pnas.1206742109} {\bibfield  {journal} {\bibinfo  {journal}
  {Proceedings of the National Academy of Sciences}\ }\textbf {\bibinfo
  {volume} {109}},\ \bibinfo {pages} {16029} (\bibinfo {year}
  {2012})}\BibitemShut {NoStop}%
\bibitem [{\citenamefont {Colombo}\ and\ \citenamefont
  {Gado}(2014)}]{colomboSelfassemblyCooperativeDynamics2014}%
  \BibitemOpen
  \bibfield  {author} {\bibinfo {author} {\bibfnamefont {J.}~\bibnamefont
  {Colombo}}\ and\ \bibinfo {author} {\bibfnamefont {E.~D.}\ \bibnamefont
  {Gado}},\ }\href {\doibase 10.1039/C4SM00219A} {\bibfield  {journal}
  {\bibinfo  {journal} {Soft Matter}\ }\textbf {\bibinfo {volume} {10}},\
  \bibinfo {pages} {4003} (\bibinfo {year} {2014})}\BibitemShut {NoStop}%
\bibitem [{\citenamefont {Gu}\ \emph {et~al.}(2018)\citenamefont {Gu},
  \citenamefont {Alt}, \citenamefont {Wang}, \citenamefont {Li}, \citenamefont
  {Willard},\ and\ \citenamefont {Johnson}}]{gu2018photoswitching}%
  \BibitemOpen
  \bibfield  {author} {\bibinfo {author} {\bibfnamefont {Y.}~\bibnamefont
  {Gu}}, \bibinfo {author} {\bibfnamefont {E.~A.}\ \bibnamefont {Alt}},
  \bibinfo {author} {\bibfnamefont {H.}~\bibnamefont {Wang}}, \bibinfo {author}
  {\bibfnamefont {X.}~\bibnamefont {Li}}, \bibinfo {author} {\bibfnamefont
  {A.~P.}\ \bibnamefont {Willard}}, \ and\ \bibinfo {author} {\bibfnamefont
  {J.~A.}\ \bibnamefont {Johnson}},\ }\href@noop {} {\bibfield  {journal}
  {\bibinfo  {journal} {Nature}\ }\textbf {\bibinfo {volume} {560}},\ \bibinfo
  {pages} {65} (\bibinfo {year} {2018})}\BibitemShut {NoStop}%
\bibitem [{\citenamefont {Li}\ \emph {et~al.}(2008)\citenamefont {Li},
  \citenamefont {Zhang}, \citenamefont {Landskron},\ and\ \citenamefont
  {Liu}}]{liSpontaneousSelfAssemblyMetal2008}%
  \BibitemOpen
  \bibfield  {author} {\bibinfo {author} {\bibfnamefont {D.}~\bibnamefont
  {Li}}, \bibinfo {author} {\bibfnamefont {J.}~\bibnamefont {Zhang}}, \bibinfo
  {author} {\bibfnamefont {K.}~\bibnamefont {Landskron}}, \ and\ \bibinfo
  {author} {\bibfnamefont {T.}~\bibnamefont {Liu}},\ }\href@noop {} {\bibfield
  {journal} {\bibinfo  {journal} {Journal of the American Chemical Society}\
  }\textbf {\bibinfo {volume} {130}},\ \bibinfo {pages} {4226} (\bibinfo {year}
  {2008})}\BibitemShut {NoStop}%
\bibitem [{\citenamefont {Bouchaud}(1992)}]{bouchaud1992weak}%
  \BibitemOpen
  \bibfield  {author} {\bibinfo {author} {\bibfnamefont {J.~P.}\ \bibnamefont
  {Bouchaud}},\ }\href {\doibase 10.1051/jp1:1992238} {\bibfield  {journal}
  {\bibinfo  {journal} {Journal de Physique I}\ }\textbf {\bibinfo {volume}
  {2}},\ \bibinfo {pages} {1705} (\bibinfo {year} {1992})}\BibitemShut
  {NoStop}%
\bibitem [{\citenamefont {Choi}\ \emph {et~al.}(2020)\citenamefont {Choi},
  \citenamefont {Holehouse},\ and\ \citenamefont
  {Pappu}}]{choiPhysicalPrinciplesUnderlying2020}%
  \BibitemOpen
  \bibfield  {author} {\bibinfo {author} {\bibfnamefont {J.-M.}\ \bibnamefont
  {Choi}}, \bibinfo {author} {\bibfnamefont {A.~S.}\ \bibnamefont {Holehouse}},
  \ and\ \bibinfo {author} {\bibfnamefont {R.~V.}\ \bibnamefont {Pappu}},\
  }\href {\doibase 10.1146/annurev-biophys-121219-081629} {\bibfield  {journal}
  {\bibinfo  {journal} {Annual Review of Biophysics}\ }\textbf {\bibinfo
  {volume} {49}},\ \bibinfo {pages} {107} (\bibinfo {year} {2020})}\BibitemShut
  {NoStop}%
\bibitem [{\citenamefont {Lieleg}\ \emph {et~al.}(2011)\citenamefont {Lieleg},
  \citenamefont {Kayser}, \citenamefont {Brambilla}, \citenamefont
  {Cipelletti},\ and\ \citenamefont {Bausch}}]{lielegSlowDynamicsInternal2011}%
  \BibitemOpen
  \bibfield  {author} {\bibinfo {author} {\bibfnamefont {O.}~\bibnamefont
  {Lieleg}}, \bibinfo {author} {\bibfnamefont {J.}~\bibnamefont {Kayser}},
  \bibinfo {author} {\bibfnamefont {G.}~\bibnamefont {Brambilla}}, \bibinfo
  {author} {\bibfnamefont {L.}~\bibnamefont {Cipelletti}}, \ and\ \bibinfo
  {author} {\bibfnamefont {A.~R.}\ \bibnamefont {Bausch}},\ }\href {\doibase
  10.1038/nmat2939} {\bibfield  {journal} {\bibinfo  {journal} {Nature
  Materials}\ }\textbf {\bibinfo {volume} {10}},\ \bibinfo {pages} {236}
  (\bibinfo {year} {2011})}\BibitemShut {NoStop}%
\bibitem [{\citenamefont {{Holten-Andersen}}\ \emph {et~al.}(2011)\citenamefont
  {{Holten-Andersen}}, \citenamefont {Harrington}, \citenamefont {Birkedal},
  \citenamefont {Lee}, \citenamefont {Messersmith}, \citenamefont {Lee},\ and\
  \citenamefont {Waite}}]{Holten-Andersen_PNAS_2010}%
  \BibitemOpen
  \bibfield  {author} {\bibinfo {author} {\bibfnamefont {N.}~\bibnamefont
  {{Holten-Andersen}}}, \bibinfo {author} {\bibfnamefont {M.~J.}\ \bibnamefont
  {Harrington}}, \bibinfo {author} {\bibfnamefont {H.}~\bibnamefont
  {Birkedal}}, \bibinfo {author} {\bibfnamefont {B.~P.}\ \bibnamefont {Lee}},
  \bibinfo {author} {\bibfnamefont {P.~B.}\ \bibnamefont {Messersmith}},
  \bibinfo {author} {\bibfnamefont {K.~Y.~C.}\ \bibnamefont {Lee}}, \ and\
  \bibinfo {author} {\bibfnamefont {J.~H.}\ \bibnamefont {Waite}},\ }\href
  {\doibase 10.1073/pnas.1015862108} {\bibfield  {journal} {\bibinfo  {journal}
  {Proceedings of the National Academy of Sciences}\ }\textbf {\bibinfo
  {volume} {108}},\ \bibinfo {pages} {2651} (\bibinfo {year}
  {2011})}\BibitemShut {NoStop}%
\bibitem [{\citenamefont {Song}\ \emph {et~al.}(2020)\citenamefont {Song},
  \citenamefont {Rizvi}, \citenamefont {Lynch}, \citenamefont {Ilavsky},
  \citenamefont {Mankus}, \citenamefont {Tracy}, \citenamefont {McKinley},\
  and\ \citenamefont
  {{Holten-Andersen}}}]{songProgrammableAnisotropyPercolation2020}%
  \BibitemOpen
  \bibfield  {author} {\bibinfo {author} {\bibfnamefont {J.}~\bibnamefont
  {Song}}, \bibinfo {author} {\bibfnamefont {M.~H.}\ \bibnamefont {Rizvi}},
  \bibinfo {author} {\bibfnamefont {B.~B.}\ \bibnamefont {Lynch}}, \bibinfo
  {author} {\bibfnamefont {J.}~\bibnamefont {Ilavsky}}, \bibinfo {author}
  {\bibfnamefont {D.}~\bibnamefont {Mankus}}, \bibinfo {author} {\bibfnamefont
  {J.~B.}\ \bibnamefont {Tracy}}, \bibinfo {author} {\bibfnamefont {G.~H.}\
  \bibnamefont {McKinley}}, \ and\ \bibinfo {author} {\bibfnamefont
  {N.}~\bibnamefont {{Holten-Andersen}}},\ }\href {\doibase
  10.1021/acsnano.0c06389} {\bibfield  {journal} {\bibinfo  {journal} {ACS
  Nano}\ }\textbf {\bibinfo {volume} {14}},\ \bibinfo {pages} {17018} (\bibinfo
  {year} {2020})}\BibitemShut {NoStop}%
\end{thebibliography}%


\begin{thebibliography}{1}

\bibitem{texierIndividualEnergyLevel2000}
Christophe Texier.
\newblock Individual energy level distributions for one-dimensional diagonal
  and off-diagonal disorder.
\newblock {\em Journal of Physics A: Mathematical and General},
  33(35):6095--6128, August 2000.

\bibitem{kampenStochasticProcessesPhysics1992}
N.~G.~Van Kampen.
\newblock {\em Stochastic {{Processes}} in {{Physics}} and {{Chemistry}}}.
\newblock {Elsevier}, November 1992.

\end{thebibliography}
\end{document}


\title{Supplementary Information \\ \large{Valence can control the nonexponential viscoelastic relaxation of multivalent reversible gels}}
\author{Hugo \surname{Le Roy} $^\dag$}
\email{h.leroy@epfl.ch}
\affiliation{Université Paris-Saclay, CNRS, LPTMS, 91405, Orsay, France}
\affiliation{Institute of Physics, \'Ecole Polytechnique F\'ed\'erale de Lausanne (EPFL), 1015 Lausanne, Switzerland}
\author{Jake Song}
\email{These authors contributed equally to this work}
\affiliation{Department of Materials Science and Engineering, Massachusetts Institute of Technology, 77 Massachusetts Avenue, Cambridge, MA 02139, USA}
\author{David Lundberg}
\affiliation{Department of Chemical Engineering, Massachusetts Institute of Technology, 77 Massachusetts Avenue, Cambridge, MA 02139, USA}
\author{Aleksandr V. Zhukhovitskiy}
\affiliation{Department of Chemistry, Massachusetts Institute of Technology, 77 Massachusetts Avenue, Cambridge, MA 02139, USA}
\affiliation{Department of Chemistry, University of North Carolina at Chapel Hill; Chapel Hill, NC 27599, USA}
\author{Jeremiah A. Johnson}
\affiliation{Department of Chemistry, Massachusetts Institute of Technology, 77 Massachusetts Avenue, Cambridge, MA 02139, USA}
\author{Gareth H. McKinley}
\affiliation{Department of Mechanical Engineering, Massachusetts Institute of Technology, 77 Massachusetts Avenue, Cambridge, MA 02139, USA}
\author{Niels Holten-Andersen}
\affiliation{Department of Materials Science and Engineering, Massachusetts Institute of Technology, 77 Massachusetts Avenue, Cambridge, MA 02139, USA}
\author{Martin Lenz}
\email{martin.lenz@universite-paris-saclay.fr}
\affiliation{Université Paris-Saclay, CNRS, LPTMS, 91405, Orsay, France}
\affiliation{PMMH, CNRS, ESPCI Paris, PSL University, Sorbonne Université, Université de Paris, F-75005, Paris, France}
\keywords{}
\maketitle
Here we present the mathematical proofs of the main results of the manuscript as well as details on the methodology used to analyze the experimental data.

\section{Distribution of superbond breaking time and derivation of $\tau_N$}
Here we show that the survival probability for the detachment of a superbond (illustrated in main text Fig.~2) containing many polymer strands ($N\rightarrow\infty$) asymptotically goes to $S(t)=e^{-t/\tau_N}$, where $\tau_N$ is given by Eq.~(4) of the main text. We first consider a general one-step process and derive the basic recursion equation used throughout the proof in Sec.~\ref{sec:BK}. We solve the recursion in Sec.~\ref{sec:sum} and express the generating function of $S(t)$ as a double sum. In Sec.~\ref{sec:Application}, we apply the resulting formula to our particular problem and take the continuum limit of the second sum. Finally, we compute both sums in the $N\rightarrow\infty$ limit in Sec.~\ref{sec:Asymptotics}. Our derivation is adapted from the calculation presented in the appendix of Ref.~\cite{texierIndividualEnergyLevel2000}.

\subsection{\label{sec:BK}Backward Kolmogorov equation for the generating function of $S(t)$}
\begin{figure}[t]
\centering
\includegraphics{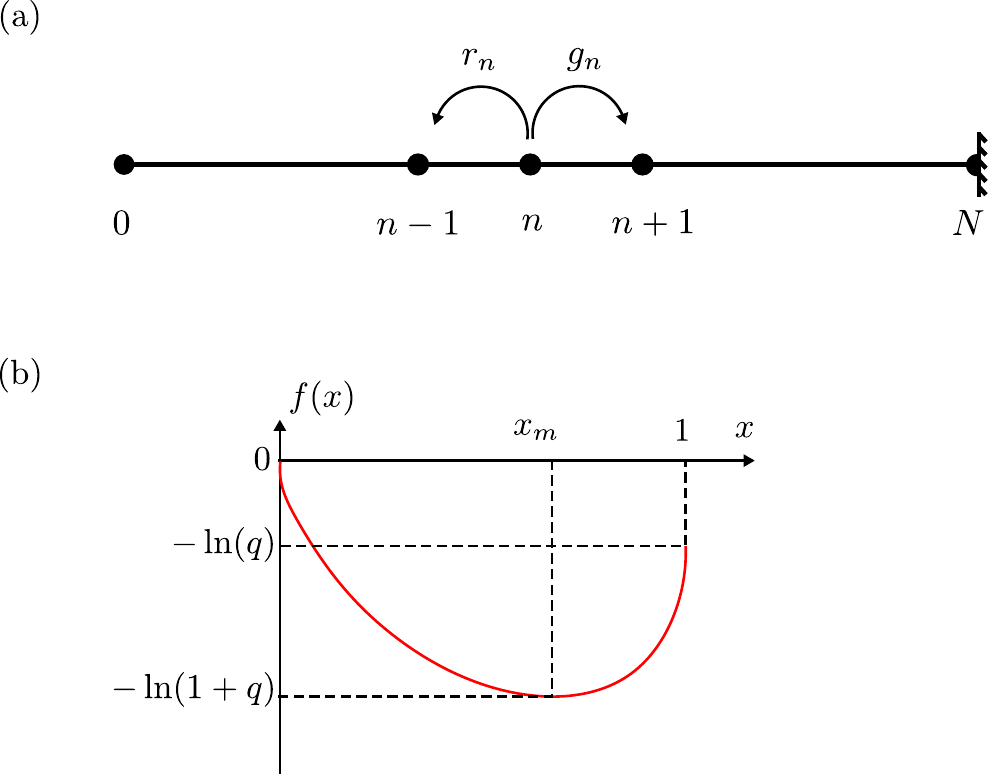}
\caption{\label{fig:Kramers_proof}Superbond detachment as a Kramers-like barrier-crossing problem. (a)~Definition of the rates of the one-step process. (b)~Profile of the pseudo-free energy defined in Eq.~\eqref{eq:free_energy}. Superbond detachment requires the system to fluctuate out of the free energy well to the $x=0$ absorbing state, with $1/N$ playing the role of a temperature.}
\end{figure}

We consider a one-step process, \emph{i.e.}, a stochastic process consisting of transitions between consecutive discrete states on a line, with transition rates $r_n$ and $g_n$ illustrated in Fig.~\ref{fig:Kramers_proof}(a). We denote the probability for the particle to be in state $k$ at time $t$ after starting in state $n$ at time $0$ by $P(k,t|n)$. We assume an absorbing boundary condition in $0$ and a reflecting boundary condition in $N$, \emph{i.e.},
\begin{equation}\label{eq:BC}
\forall n\in[1..N]
\quad
P(0,t|n)=0,
\qquad
r_N=0.
\end{equation}
The backward Kolmogorov equation for our process reads~\cite{kampenStochasticProcessesPhysics1992}
\begin{equation}
\frac{\text{d}P}{\text{d}t}(k,t|n)=
g_n[P(k,t|n+1)-P(k,t|n)]
-r_n[P(k,t|n)-P(k,t|n-1)].\label{Eq:BK_for_p}
\end{equation}
We define the survival probability and its generating function (Laplace transform), respectively as
\begin{equation}
S_n(t)=\sum_{k=1}^NP(k,t|n),
\quad
h_n(\alpha)=\int_0^{+\infty}S_n(t)e^{-\alpha t}\,\text{d}t.
\end{equation}
Inserting these definitions into Eq.~\eqref{Eq:BK_for_p} yields
\begin{equation}
\alpha h_n(\alpha)-1=g_n[h_{n+1}(\alpha)-h_n(\alpha)]
-r_n[h_{n}(\alpha)-h_{n-1}(\alpha)], \label{Eq:BK_for_h}
\end{equation}
which we endeavor to solve for $h_n(\alpha)$ in the following.

\subsection{\label{sec:sum}Sum equation for the generating function}
We define a rescaled current between sites  $n-1$ and $n$
\begin{equation}\label{eq:Delta_def}
\Delta_n =
\begin{cases}
r_n \left(\prod_{i=n}^{N-1}\frac{r_{i+1}}{g_i}\right)[h_n-h_{n-1}] & \text{for }n<N\\
r_N [h_N-h_{N-1}]& \text{for }n=N
\end{cases}.
\end{equation}
This allows us to turn the two-step recursion of Eq.~\eqref{Eq:BK_for_h} into one with only one step:
\begin{equation}
\Delta_n =
\begin{cases}
\Delta_{n+1} + \left(\prod_{i=n}^{N-1}\frac{r_{i+1}}{g_i}\right)[1-\alpha h_n] & \text{for }n<N\\
1-\alpha h_N& \text{for }n=N
\end{cases},
\end{equation}
which can easily be summed as
\begin{equation}\label{eq:Delta_solution}
\Delta_n=\left[\sum_{j=n}^{N-1}
\left(\prod_{i=n}^{N-1}\frac{r_{i+1}}{g_i}\right)
(1-\alpha h_j)
\right]+1-\alpha h_N.
\end{equation}
We now invert Eq.~\eqref{eq:Delta_def} and use Eq.~\eqref{eq:Delta_solution} to express the finite difference $(h_n-h_{n-1})$. We further use the property that $h_m=h_0+\sum_{n=1}^m(h_n-h_{n-1})$ and recognize that $h_0=0$ due to Eq.~\eqref{eq:BC} to obtain
\begin{widetext}
\begin{equation}\label{eq:sum_equation}
h_m=\sum_{n=1}^m\frac{1}{r_N}\left(\prod_{i=n}^{N-1}\frac{g_i}{r_{i+1}}\right)
\left\lbrace
\left[\sum_{j=n}^{N-1}
\left(\prod_{i=n}^{N-1}\frac{r_{i+1}}{g_i}\right)
(1-\alpha h_j)
\right]+1-\alpha h_N
\right\rbrace.
\end{equation}
\end{widetext}

\subsection{\label{sec:Application}Application and continuum limit}
Using the mean detachment time of a polymer strand (denoted as $\omega_-$ in the main text) as our unit of time and defining $q=\omega_+/\omega_-$, the model of the main text implies
\begin{equation}
\forall n\in[1..N]\quad
r_n=n,
\quad
g_n=(N-n)q,
\end{equation}
which we insert into Eq.~\eqref{eq:sum_equation} to obtain
\begin{equation}\label{eq:applied_sum}
h_n=\sum_{j=1}^n\frac{
1
}
{
j\binom{N}{j}q^j
}\sum_{i=j}^N\binom{N}{i}q^i(1-\alpha h_i).
\end{equation}
In Eq.~\eqref{eq:applied_sum}, the outermost sum is dominated by the very small values of $j$ in the limit $N\rightarrow\infty$. We thus need only consider small values of $j$ when computing the innermost sum, which happens to be dominated by a value of $i$ far from the edges of the $[1..N]$ interval. We can thus take its continuum limit. Using Stirling's formula, we obtain
\begin{equation}\label{eq:integral}
h_n\underset{N\rightarrow\infty}{\sim}\sum_{j=1}^n\frac{1
}
{
j\binom{N}{j}q^j
}
\int_{0}^1\sqrt{\frac{N}{2\pi x(1-x)}}e^{-Nf(x)}[1-\alpha h(x,\alpha)]\,\text{d}x,
\end{equation}
where we have defined the continuum version of our generating function though $h(x,\alpha)=h_{Nx}(\alpha)$, as well as the pseudo free energy of the system
\begin{equation}
f(x)=x\ln x+(1-x)\ln(1-x)-x\ln q.
\end{equation}
This free energy has a single minimum in $x_m=q/(1+q)$ with a locally parabolic structure given by
\begin{equation}\label{eq:free_energy}
f(x)=-\ln(1+q)+\frac{(1+q)^2}{2q}(x-x_m)^2+\mathcal{O}(x-x_m)^3,
\end{equation}
which we illustrate in Fig.~\ref{fig:Kramers_proof}(b).
The problem at hand is exactly analogous to a Kramers escape problem from the bottom of this minimum to the $n=0$ boundary condition, with $N\rightarrow\infty$ playing the role of the low-temperature limit.

\subsection{\label{sec:Asymptotics}Asymptotic simplifications}
Using the Kramers analogy to our advantage, we compute the integral of Eq.~\eqref{eq:integral} using a saddle-point approximation. We thus find that for any $x\in]0,1[$:
\begin{equation}\label{eq:h_1}
h(x,\alpha)\underset{N\rightarrow\infty}\sim
(1+q)^N[1-\alpha h(x_m,\alpha)]\sum_{j=1}^{Nx}\frac{q^{-j}}{j\binom{N}{j}}.
\end{equation}
Using Stirling's formula for small values of $j$ reveals that the argument of the sum in Eq.~\eqref{eq:h_1} goes as $(j-1)!\times (Nq)^{-j}$. Therefore, the terms of the sum are simply the terms in an expansion in powers of $N$. We keep only the lowest-order term to find
\begin{equation}\label{eq:h_x_xm}
\forall x\in]0,1[\quad
h(x,\alpha)\underset{N\rightarrow\infty}\sim
\tau_N[1-\alpha h(x_m,\alpha)].
\end{equation}
where
\begin{equation}\label{eq:dimensionless_tau_N}
\tau_N=\frac{(1+q)^N}{Nq}
\end{equation}
is the dimensionless version of the mean first-passage time presented in Eq.~(4) of the main text.

Setting $x=x_m$, Eq.~\eqref{eq:h_x_xm} implies
\begin{equation}
h(x_m,\alpha)\underset{N\rightarrow\infty}\sim\frac{1}{\alpha+\tau_N^{-1}}
\quad\Leftrightarrow\quad
S_{Nx_m}(t)\underset{N\rightarrow\infty}\sim e^{-t/\tau_N}.
\end{equation}
Finally, using Eq.~\eqref{eq:h_x_xm} again yields
\begin{equation}
\forall x\in]0,1[\quad
S_{Nx}(t)\underset{N\rightarrow\infty}\sim-\tau_N\frac{\text{d} S_{Nx_m}}{\text{d}t}(t)=e^{-t/\tau_N},
\end{equation}
which is the exponential distribution presented in the main text.
\section{Link between $\alpha$ and $N_\text{sat}/ \bar{N}$}
\begin{figure}[t]
    \centering
    \includegraphics[height=15cm]{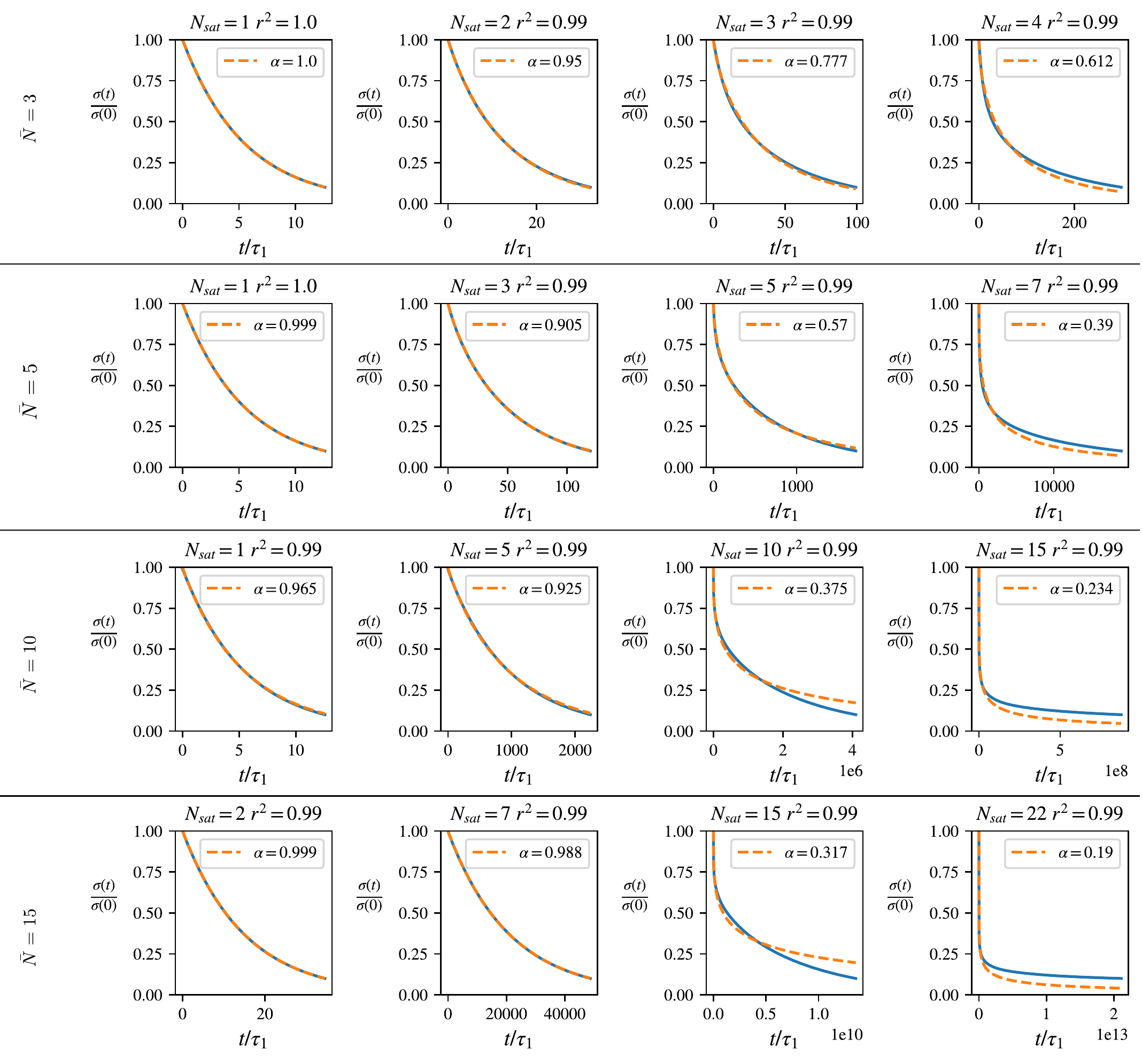}
    \caption{Illustration of the similarity of our modeled stress response function with a stretched exponential.
    We plot the relaxation modulus computed using Eq.~(6) of the main text for $\bar{N} \in [3,5,10,15]$. For each value of $\bar{N}$, we plot four values of $N_\text{sat}$, namely $N_\text{sat}=0.1\bar{N}$, $0.5\bar{N}$, $\bar{N}$ and $1.5\bar{N}$, $p_\text{off} = 0.2$. Each plot also mentions the value of the fitted stretch exponent $\alpha$ and the correlation coefficient $r^2$.
}
    \label{fitsstretch}
\end{figure}

\begin{figure}[t]
\centering
\includegraphics[width=15cm]{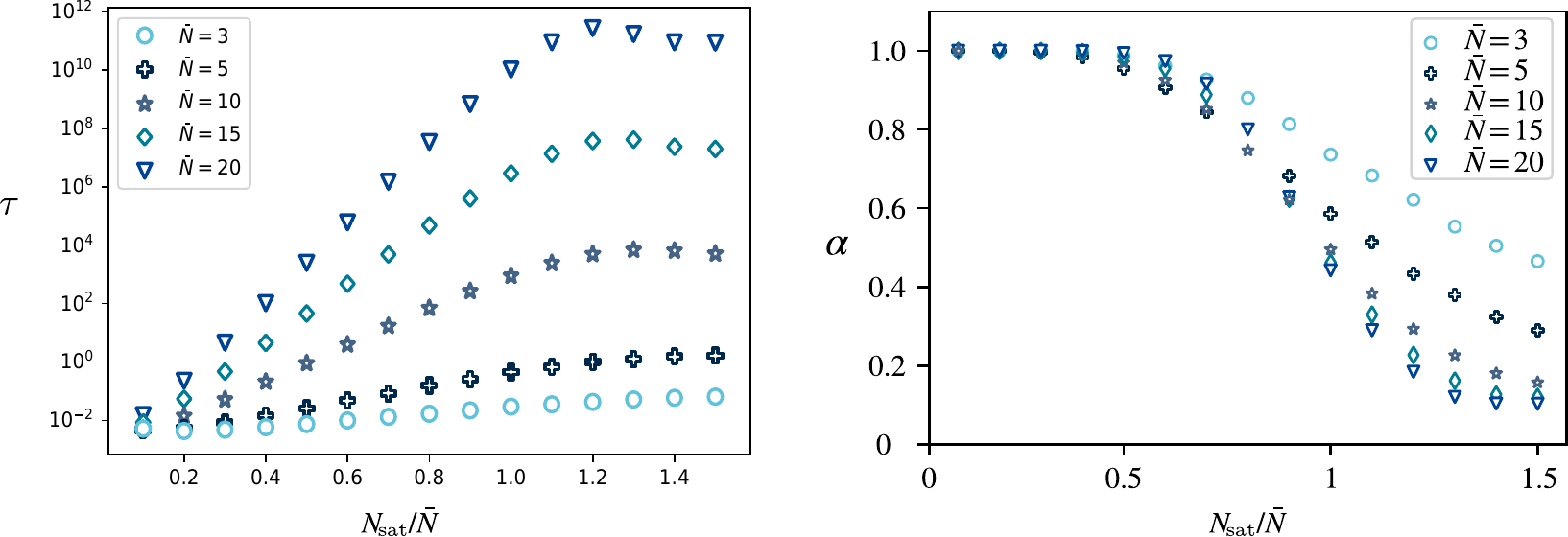}
\caption{\label{fig:taualphaNcut} Best fit values of the stretched exponential parameters $\tau$ and $\alpha$ for a range of values of $\bar{N}$ and $N_\text{sat}$. The right-hand-side panel is identical to Fig.~4 of the main text.
}
\end{figure}

Here establish the connection between the stretch exponent $\alpha$ and the values of $N_\text{sat}/\bar{N}$ shown in Fig.~4 of the main text. To mimic the observation of an experimental step strain over a finite time window, we focus our attention on the time interval between $t=0$ and $t=\tau_{90}$, where $\tau_{90}$ is the time required to relax $90\%$ of the stress, \emph{i.e.}, $\sigma(\tau_{90})=0.1\times \sigma(0)$. We plot the relaxation curve given by Eq.~(6) of the main text over this time window, then perform a least-squares fit using a stretched exponential [Eq.~(1) of the main text] with $\alpha$ and $\tau$ as fitting parameters. As shown in Fig.~\ref{fitsstretch}, the agreement is excellent for a large majority of the parameters used. The corresponding value of the fitting parameters ($\tau$ and $\alpha$) for a broader variety of $\bar{N}$ and $N_\text{sat}$ is also provided in Fig.~\ref{fig:taualphaNcut}. This suggests that experimental curves that are well fitted by a stretched exponential could be equally well described by our model. 

\newpage

\section{Time-temperature collapse}
Here we describe the procedure used to determine the binding energy $\Delta E$ in the experimental systems discussed in the main text. Equation~(4) of the main text implies that the temperature dependence of the stress response function can be eliminated by expressing it as a function of the rescaled time $\tilde{t}=te^{\beta \Delta E}$. This should cause the relaxation curves of a given system at different temperatures to collapse.

For each type of ligand, we have 5 datasets showing the stress relaxation function as a function of time at each different temperature $\{T^{(\alpha)}\}_{\alpha\in [0,4]} = \{25^{\circ}\text{C},\,35^{\circ}\text{C},\,45^{\circ}\text{C},\,55^{\circ}\text{C},\,65^{\circ}\text{C}\}$. To enable the comparison between time-rescaled datasets, we first define an interpolating function for the stress relaxation function at each temperature used.
We thus compute the set of interpolating coefficients $\left\{p^{(\alpha)}_k\right\}_k$ by perform a least-square fit of the following rational function
\begin{equation}
    P^{(\alpha)}(t) = \sum_{k = -3}^{10} p^{(\alpha)}_k t^k,
\end{equation}
to the datapoints $\left\{t^{(\alpha)}_i,\frac{\sigma^{(\alpha)}(t_i)}{\sigma^{(\alpha)}(0)}\right\}_i$. We furthermore define the interval of definition of $P^{(\alpha)}(x)$ as the range over which data is available, \emph{i.e.}, $I_{P^{(\alpha)}}=\left[0,\underset{i}{\max}\,t^{(\alpha)}_i\right]$.

\begin{figure}[t]
\centering
\includegraphics{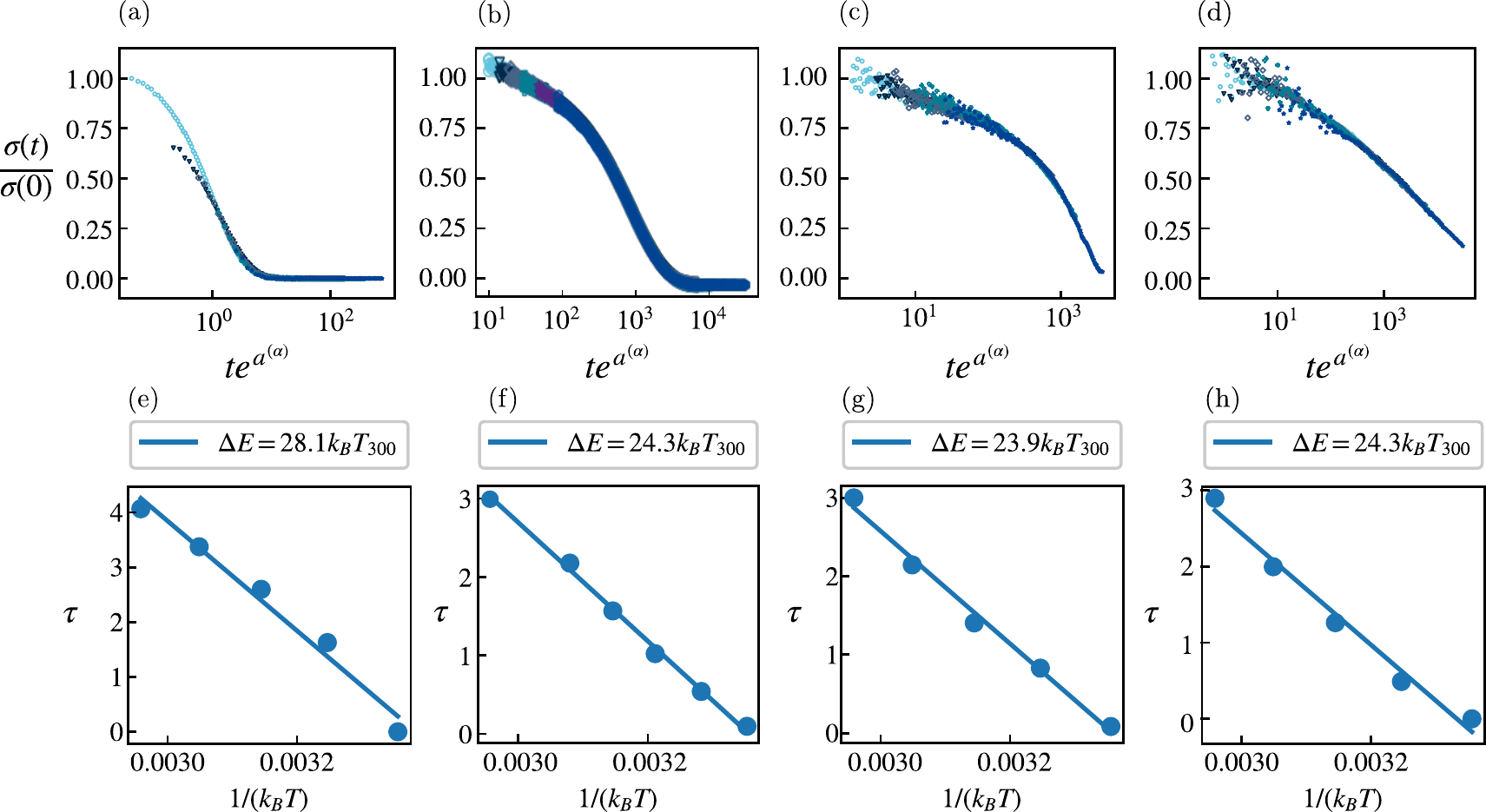}
\caption{\label{fig:collapse}\textbf{Collapse of the relaxation modulus: } (a-c) Collapsed relaxation modulus of $\text{Fe}^{3+}$ ions, Pd$_{12}$L$_{24}$ nanocages and nanoparticles respectively after the rescaling of the time for an optimized collapse
. The curves are represented on a log-lin coordinate system, but the collapsing procedure is performed on a lin-lin scale 
. (d-f) corresponding rescaling parameters as a function of $1/(k_BT)$ the inverse temperature. The slope of the line is $-\Delta E$ and the legend gives the value of $\Delta E$ in $k_B T$ unit at $300K$.}
\end{figure}

We then perform the collapse of the $\{T^{(\alpha)}\}_{\alpha\in [1,4]}$  interpolated curves onto the $T^{(0)}$ curve. To this effect we define the set of rescaling coefficients $\{a^{(\alpha)}\}_{\alpha\in [1,4]}$ and performs a separate time rescaling for each temperature: $\tilde{t}= te^{a^{(\alpha)}}$. For each $\alpha\in[1,4]$, we optimise the semidistance
\begin{equation}
D(P,Q) = \int_{I_Q \cap I_P}[P(t)-Q(t)]^2 \text{d}t,
\label{eq:distance}
\end{equation}
between the functions $t\rightarrow P^{(0)}(t)$ and $t\rightarrow P^{(\alpha)}(te^{a^{(\alpha)}})$ with respect to $a^{(\alpha)}$. The resulting collapsed curves are shown in Fig.~\ref{fig:collapse}~(a,b,c). The optimal rescaling coefficients are plotted as a function of the inverse temperature ${1}/{k_BT}$ in Fig.~\ref{fig:collapse}~(d,e,f). Consistent with the time-temperature collapse hypothesis, this dependence is affine, and we use the slope of the best fitting line as our value of the binding energy $\Delta E$.

\section{Fit of the stress relaxation function to our theoretical prediction}
\begin{figure}[t]
    \centering
    \includegraphics[width=18cm]{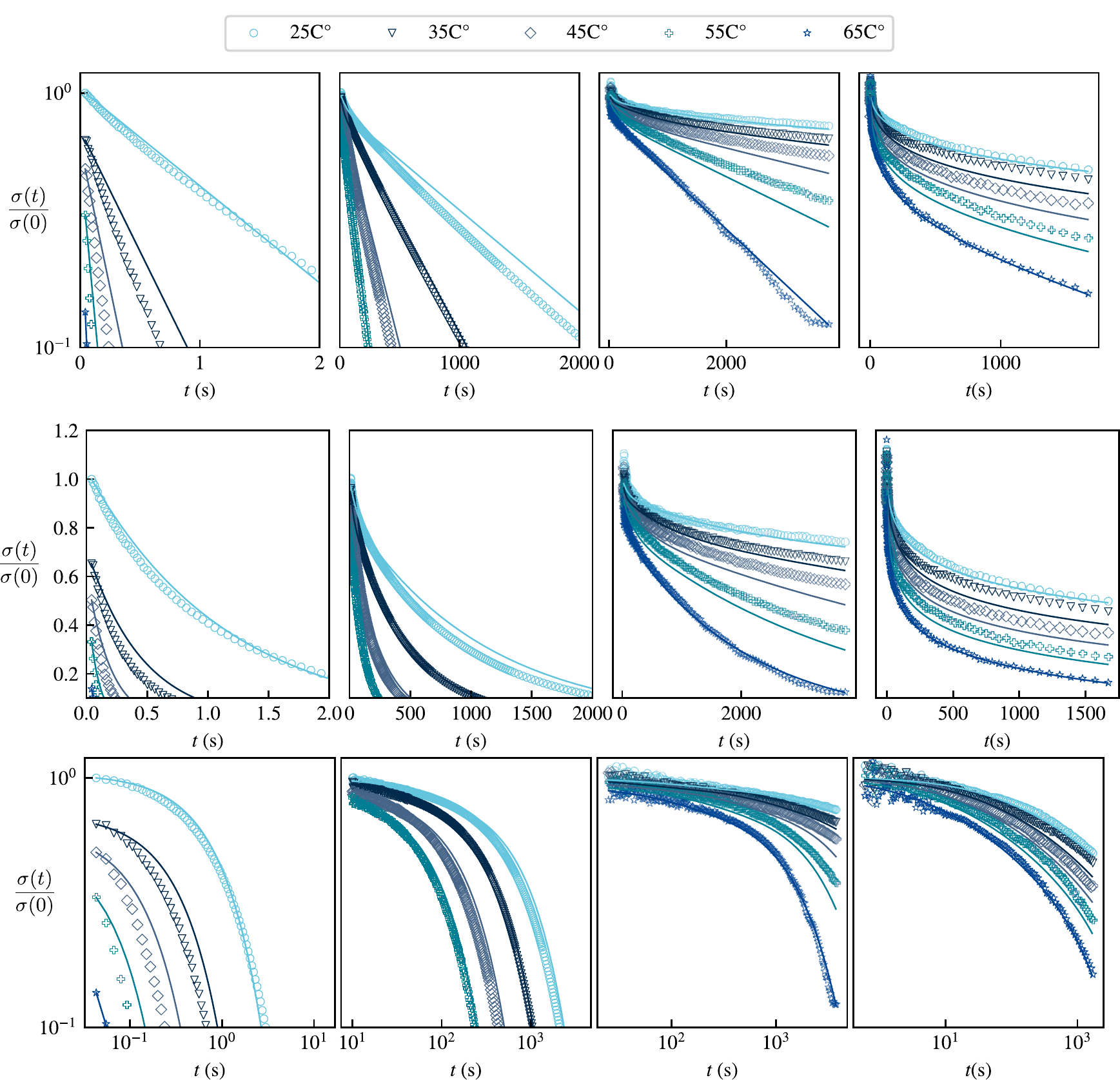}
    \caption{\label{fig:Fits}Fits of the experimental curves : respectively lin-lin and lin-log and log-log representation of Fig.~5 in the main text.}
\end{figure}

In the main text, we fit the experimental curves with the stress relaxation function predicted by our model. We then represent them on a log-lin scale to allow the simultaneous visualization of short and long time scales. To demonstrate the robustness of our fits, in Fig.~\ref{fig:Fits} we replot these curves in a lin-lin-scale, as well as a lin-log scale that emphasizes intervals of exponential relaxation as straight lines.

\section{Rationalization of the Poisson distribution of the superbond size $p(N)$}
The polymers used in our experiments are 4-arms polyethylene glycol (PEG). At the end of each arm is a nitrocatechol ligand that allows crosslinker binding. In our model, we assume that the ends of a polymer are always attached to a ligand. For this reason, the diffusion of such a polymer over a distance comparable to the polymer size occurs on a time scale comparable to the time required to rearrange the bonds between crosslinkers, which corresponds to the time required for the relaxation of the stress in the system.
Let us consider that the 4-arm PEG are able to diffuse over a volume $v$ during the time of the experiment.
We model the spreading of the polymers in the system by discretizing the system into small boxes of volume $v$ between which no polymer exchange occurs over the duration of the experiment. As a result the distribution of the polymers over the boxes is due to the initial preparation of the system. We assume that this processes places each polymer in a random box with equal probability. As a result, the probability that a specific box contains $n$ polymers is given by a Poisson distribution:
\begin{equation}
P(n) = e^{-\rho_\text{PEG}v}\frac{(\rho_\text{PEG}v)^{n}}{n!},
\label{PoissonBox}
\end{equation}
where $\rho_\text{PEG}$ is the average concentration of PEG in the system, and $v\rho_\text{PEG}$ is the mean (over the system) number of $PEG$ in a box of volume $v$. Equation~\eqref{PoissonBox} is the basis for Eq.~(5) of the main text.
\section{Experimental fit using alternative distributions of $N$}
\begin{figure}
    \centering
    \includegraphics[width=17cm]{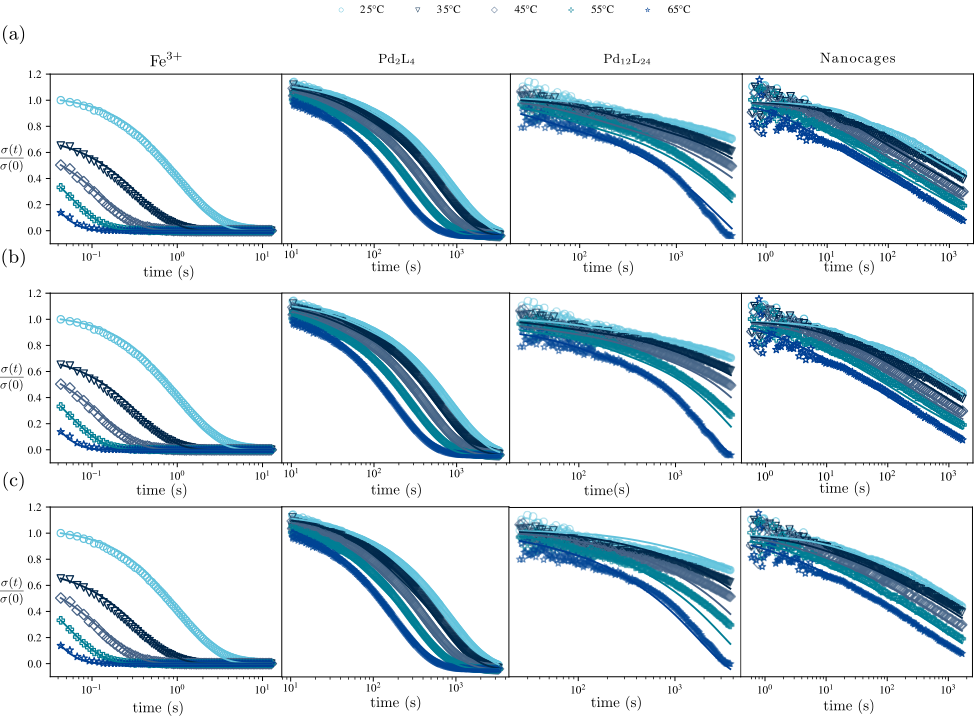}
    \caption{Fit of the experimental data using alternative probability distribution of $N$. Respectively Fe$^{3+}$, nanoparticles and nanocage relaxation curves fitted using (a) a uniform distribution, (b) a triangular distribution and (c) using a linear distribution.}
    \label{fig:alternative_fit}
\end{figure}

To demonstrate the robustness of our model, we perform the fit of the experimental data using probability distribution different from the Poisson distribution of Eq.~(5) of the main text.
For this, we first define three new distributions with the same mean (and when possible the same variance) than the one of Eq.~(5) of the main text and keep the saturation value identical to the one in the main text. We first use a rectangular distribution :
\begin{subequations}    
\begin{align}
    \text{if }\bar{N}-\sqrt{3\bar{N}}<0 &\quad
    p_\text{rectangular}(N) = 
    \begin{cases}
    \frac{1}{\bar{N} +\sqrt{3\bar{N}}} & \text{for } N \in [0,\bar{N}+\sqrt{3\bar{N}}] \\
    0 & \text{otherwise}
    \end{cases}\\
    \text{if }\bar{N}-\sqrt{3\bar{N}}>0 &\quad
    p_\text{rectangular}(N) = 
    \begin{cases}
    \frac{1}{2\sqrt{3\bar{N}}} & \text{for } N \in [\bar{N}-\sqrt{3\bar{N}},\bar{N}+\sqrt{3\bar{N}}] \\
    0 & \text{otherwise}
    \end{cases},
\end{align}
\end{subequations}
a triangular distribution:
\begin{equation}
    p_\text{triangular}(N) = 
    \begin{cases}
    \frac{N}{\bar{N}^2} & \text{for } N \in [0,\bar{N}] \\
    \frac{2}{\bar{N}}-\frac{N}{\bar{N}^2} & \text{for } N \in [\bar{N},2\bar{N}]\\
    0 & \text{otherwise}
    \end{cases}
    ,
\end{equation}
and a linear distribution:
\begin{equation}
    p_\text{linear}(N) =
    \begin{cases}
    \frac{8 N}{9\bar{N}^2} & \text{if } N\in [0,3\bar{N}/2] \\
    0 & \text{otherwise}
    \end{cases}.
\end{equation}
In each of these cases, we introduce a truncation at the maximum superbond size in the same way as in Eqs.~(5-6) of the main text (with discrete sums replaced by integrals). The resulting fits are displayed in Fig.~\ref{fig:alternative_fit}, and the corresponding value of the fitting parameters are given in Table~\ref{tab:parameters}. The fitting curves and fitting parameters remain close to the ones obtained in the main text, implying that our specific choice of a Poisson distribution of superbond sizes is not critical for the validity of our results.
\begin{table}
\centering
\setlength{\tabcolsep}{6pt} 
\renewcommand{\arraystretch}{1.2}
\begin{tabular}{|c|c|c|c|c|c|}
\hline
    & parameter & Fe$^{3+}$ & Pd$_2$L$_4$ &Pd$_{12}$L$_{24}$ & nanoparticles \\
\hline
     \multirow{3}{*}{\rotatebox[origin=c]{90}{uniform}} & 
     $\mathbf{p}_\mathbf{off}$ & 0.08 & 0.07 & 0.06 & 0.3\\
     &$\mathbf{\tau_1}$(s) & 0.86 & 122 & 41 & 11\\
     &$\mathbf{\bar{N}}$ & 1& 2 & 4 & 7\\
\hline
\hline
    \multirow{3}{*}{\rotatebox[origin=c]{90}{triangular}} &
     $\mathbf{p}_\mathbf{off}$ & 0.09 & 0.02 &0.03 & 0.5\\
     & $\mathbf{\tau_1}$(s) & 0.86& 4 & 83 & 13\\
     & $\mathbf{\bar{N}}$ & 1 & 3 & 4 & 5\\
\hline
\hline
\multirow{3}{*}{\rotatebox[origin=c]{90}{linear}} &
     $\mathbf{p}_\mathbf{off}$ &0.08& 0.04&0.07&0.24\\
     &$\mathbf{\tau_1}$(s) & 0.86&62 & 3.9 $10^2$ & 5.7\\
     &$\mathbf{\bar{N}}$ & 1& 2 & 3.3 & 6.3\\
\hline
\end{tabular}
\caption{\label{tab:parameters}Value of the fitting parameters for the three alternative distributions of $N$ indicated in the left column.}
\end{table}

\section{\label{sec:lognormal}Derivation of the log-normal probability distribution.}
Here we derive Eq.~(7) of the main text. Considering the Eq.~(5) of the main text without the cutoff due to the valence, the distribution of single strands in a superbond becomes a simple Poisson distribution of mean $\bar{N}$. A random number drawn from such a distribution can also be considered as the sum of $\bar{N}$ independent variables drawn from a Poisson distribution of mean $1$. According to the central limit theorem, the probability distribution of the sum of $\bar{N}$ independent random numbers converges to a normal distribution of mean $\bar{N}$ and variance $\sqrt{\bar{N}}$ for $\bar{N}\gg 1$. 
Moreover, as the main dependence of $\tau_N$ on $N$ is exponential, in the large-$\bar{N}$ limit replacing the factor of $N$ preceding $p_\text{off}^N$ by the typical value $\bar{N}$ induces only a small (logarithmic) error. We thus approximate:
$\tau_N/\tau_1 \sim \exp[N\log(p_\text{off})]/\bar{N}$, and write:
\begin{equation}
    N = - \frac{\ln( \bar{N}\tau_N/(\tau_1 p_\text{off}))}{\ln(p_\text{off})}.
\end{equation}
Treating $N$ as a continuous variable when computing the stress relaxation function allows us to change variable:
\begin{equation}
    \begin{aligned}
        \frac{G(t)}{G(0)} &= \sum_N p(N) \exp(-t/\tau_N) \\
        &\approx \int_N p(N) \exp(-t/\tau_N) dN \\
        & = \int_\tau p(\tau) \exp(-t/\tau) d\tau.
    \end{aligned}
\end{equation}
Where we used in the last line the identity:
\begin{equation}
    p(\tau){d\tau} = p(N) dN,
\end{equation}
Substituting $N$ to $\tau$ in the approximate normal distribution of $N$ finally gives a log-normal distribution of relaxation time presented in Eq.~(7) of the main text.

\section{\label{sec:power}Derivation of the power law relaxation} 
\begin{figure}[t]
    \centering
    \includegraphics[width=15cm]{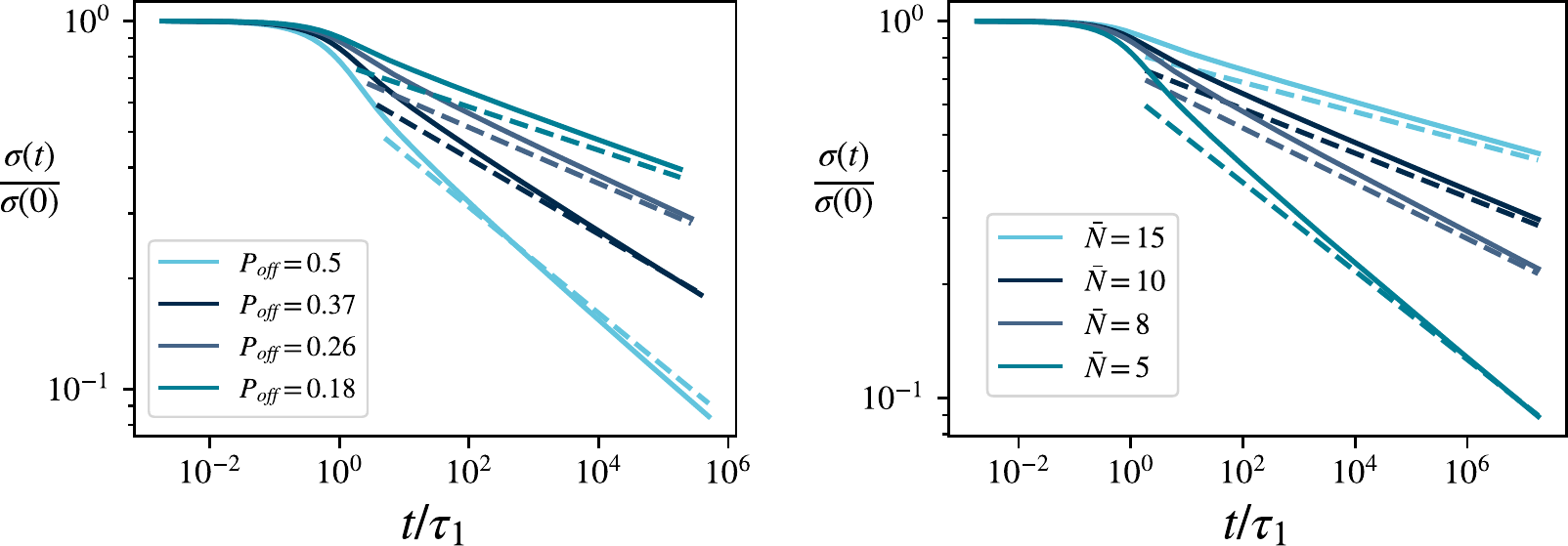}
    \caption{Comparison between exact relaxation modulus of Eq.~\eqref{eq:sup:stress} (solid lines) and the approximate expression of Eq.~\eqref{eq:power_law} (dashed lines).}
    \label{fig:supp:power_law}
\end{figure}

Here we derive Eq.~(8) of the main text.
As discussed in the main text, substituting the superbond size distribution Eq.~(5) of the main text for an exponential distribution
\begin{equation}\label{eq:exp_distrib}
    p(N) = \left(1-e^{-1/\bar{N}}\right)e^{-N/\bar{N}}
\end{equation}
yields a power-law relaxation regime provided that $\bar{N}\gg 1$, as shown in Fig.~\ref{fig:supp:power_law}. Here we compute the value of the relaxation exponent.

Since Eq.~\eqref{eq:exp_distrib} does not saturate at a finite $N=N_\text{sat}$, Eq.~(6) of the main text becomes
\begin{equation}
\label{eq:sup:stress}
    \frac{\sigma(t)}{\sigma(0)}=\sum_{N=1}^{+\infty} \frac{p(N)}{1-p(0)} e^{-t/\tau_N}
    \qquad
    \text{with}
    \qquad
    \tau_N = \frac{\tau_0 e^{\beta \Delta E}}{N p_\text{off}^N}.
\end{equation}
We employ the same approximation as in Sec.~\ref{sec:lognormal}, which captures the dominant exponential relationship between $\tau$ and $N$:
\begin{equation}\label{eq:approx_tau_N}
\tau_N \simeq \frac{\tau_0 e^{\beta \Delta E}}{\bar{N} p_\text{off}^N}.
\end{equation}
We also take the continuum limit of the sum of Eq.~\eqref{eq:sup:stress} as is appropriate for large $\bar{N}$. Defining the dimensionless time $\tilde{t}=t\bar{N}/\tau_0e^{\beta\Delta E}$, this yields
\begin{equation}
    \frac{\sigma(t)}{\sigma(0)}\underset{\bar{N}\gg 1}{\sim}\int_0^{+\infty}p(N)e^{-\tilde{t}p_\text{off}^N}\,\text{d}N.
\end{equation}
We next change our integration variable to $\tilde{\tau}=p_\text{off}^N$ to find
\begin{equation}\label{eq:power_law}
    \frac{\sigma(t)}{\sigma(0)}\underset{\bar{N}\gg 1}{\sim}\int_1^{+\infty}\gamma \tilde{\tau}^{-(1+\gamma)}e^{-\tilde{t}/\tilde{\tau}}\,\text{d}\tilde{\tau}
    \underset{\bar{N}\gg 1, \tilde{t}\gg 1}{\sim}
    \Gamma(1+\gamma)\tilde{t}^{-\gamma}
    ,
    \qquad
    \text{where}
    \qquad
    \gamma=-\frac{1}{\bar{N}\ln p_\text{off}}>0
\end{equation}
and where $\Gamma$ denotes the gamma function. Equation~\eqref{eq:power_law} implies the power law presented in Eq.~(8) of the main text, and its accuracy at long times is confirmed by the plots of Fig.~\eqref{fig:supp:power_law}. As discussed in the main text in relation to trap models, here an exponential distribution of $N$ combined with an exponential dependence of the relaxation time on $N$ [Eq.~\eqref{eq:sup:stress}] result in a power law distribution of the relaxation times. This distribution is apparent in the integral on the left of Eq.~\eqref{eq:power_law}, and eventually results in the power law relaxation. Note that the approximation of Eq.~\eqref{eq:approx_tau_N} leads us to ignore a possible logarithmic dependence of $\sigma(t)\times t^\gamma$ on $t$, hence the small mismatch between the curves of Fig.~\ref{fig:supp:power_law}.

\section{Scaling regimes for the complex modulus} 
Here we derive Eq.~(9) of the main text, the associated prefactors and its extension to $\gamma\geqslant 1$.
In the linear response regime, the Fourier transform of the stress is related to that of the strain $\epsilon$ through the material's complex modulus $G$:
\begin{equation}
    \label{eq:linear_response}
    \sigma(\omega) = G(\omega) \epsilon(\omega).
\end{equation}
Denoting the Heaviside step function by $H$, we consider the response to a step strain $\epsilon(t)=\epsilon_0 H(t)$ and thus obtain $\sigma(\omega)$ by Fourier transforming Eq.~\eqref{eq:sup:stress}. Equation~\eqref{eq:linear_response} then yields
\begin{equation}\label{eq:intermediate_G}
    \int_{0}^{+\infty} e^{-i\omega t} {\sigma(0)}\sum_{N=1}^{N_\text{sat}} \frac{p(N)}{{1-p(0)}} e^{-t/\tau_N}  \,\text{d}t
    =
    G(\omega) \int_{-\infty}^{+\infty} e^{-i\omega t} \epsilon_0 H(t)  \,\text{d}t
\end{equation}
where the bounds of the left-hand-side integral stem from the implicit assumption that $\sigma(t<0)=0$ in Eq.~\eqref{eq:sup:stress}. We compute both integrals in Eq.~\eqref{eq:intermediate_G} to find
\begin{equation}\label{eq:full_G}
    \tilde{G}(\omega) = \sum_{N=1}^{N_\text{sat}} \frac{p(N)}{1-p(0)} \frac{i\omega \tau_N}{1+i\omega \tau_N},
\end{equation}
where $\tilde{G}$ is the dimensionless modulus obtained by normalizing $G$ by the high-frequency elastic plateau $\sigma(0)/\epsilon_0$.

In the following we consider a generalization of Eq.~\eqref{eq:exp_distrib} where $p(N)\propto \exp(-N/\bar{N})$ for $N\leq N_\text{sat}$ and $p(N)=0$ for $N> N_\text{sat}$. We analyze the scaling behavior of the storage modulus $G'(\omega) = \Re[G(\omega)]$ and the loss modulus $G''(\omega) = \Im[G(\omega)]$ computed from Eq.~\eqref{eq:full_G}.

In the high-frequency regime $\omega\gg\tau_1^{-1}$, the system displays a Maxwell-like rheology:
\begin{subequations}\label{eq:omega_high}
\begin{align}
    \tilde{G}'(\omega) & \underset{\tau_1^{-1}\ll\omega}{\sim} 1
    \\
    \tilde{G}''(\omega) & \underset{\tau_1^{-1}\ll\omega}{\sim} \frac{e^{-1/\gamma\bar{N}}(1-e^{-1/\bar{N}})}{[1-e^{-(1+\gamma^{-1})/\bar{N}}]^2}\frac{1}{\omega\tau_1}.
\end{align}
\end{subequations}

We now consider the intermediate frequency regime $\tau_{N_\text{sat}}^{-1}\ll\omega\ll\tau_1^{-1}$ in the case $N_\text{sat}\gg 1$. Provided we also assume $1\ll\bar{N}\ll N_\text{sat}$, the approximate power law response of Eq.~\eqref{eq:power_law} applies and we obtain
\begin{subequations}\label{eq:omega_power}
\begin{align}
    \tilde{G}'(\omega) & \underset{\tau_{N_\text{sat}}^{-1}\ll\omega\ll\tau_1^{-1}}{\sim} 
    \begin{cases}
    \frac{\pi\gamma/2}{\sin(\pi\gamma/2)}e^{-1/\bar{N}} \left(\frac{\omega\tau_1}{\bar{N}}\right)^\gamma & \text{if }\gamma<2\\
    \frac{\gamma}{\gamma-2}e^{-2/\gamma\bar{N}}\left(\frac{\omega\tau_1}{\bar{N}}\right)^2 & \text{if }\gamma>2\\
    \end{cases}
    \\
    \tilde{G}''(\omega) & \underset{\tau_{N_\text{sat}}^{-1}\ll\omega\ll\tau_1^{-1}}{\sim}
    \begin{cases}
    \frac{\pi\gamma/2}{\cos(\pi\gamma/2)}e^{-1/\bar{N}} \left(\frac{\omega\tau_1}{\bar{N}}\right)^\gamma & \text{if }\gamma<1\\
    \frac{\gamma}{\gamma-1}e^{-1/\gamma\bar{N}}\left(\frac{\omega\tau_1}{\bar{N}}\right) & \text{if }\gamma>1
    \end{cases}.
\end{align}
\end{subequations}

\begin{figure}[t]
    \centering
    \includegraphics[width=16cm]{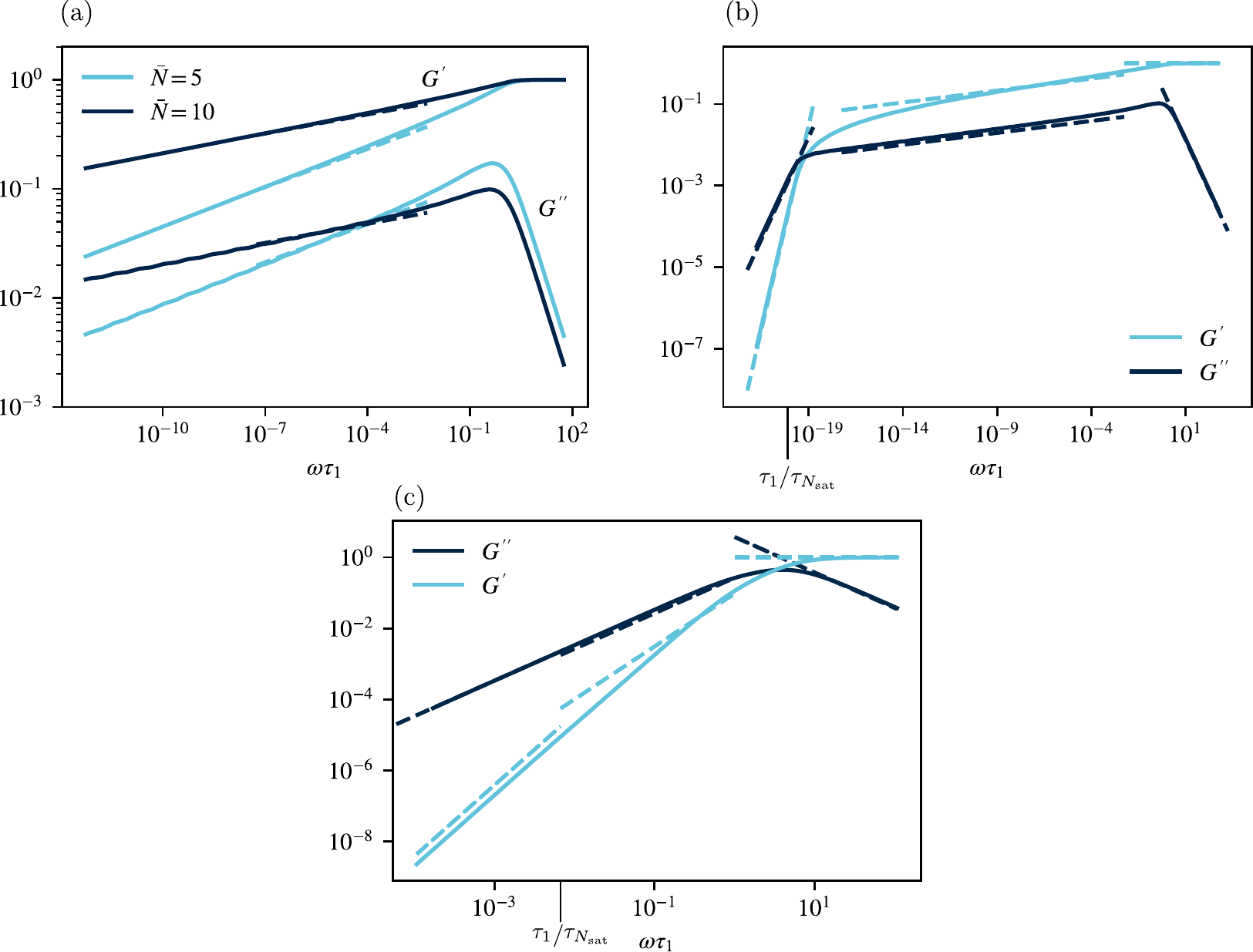}
    \caption{
    Comparison between the storage and loss moduli computed from the exact expression Eq.~\eqref{eq:full_G} (solid lines) and the asymptotic expressions of Eqs.~(\ref{eq:omega_high}-\ref{eq:omega_low}) (dashed lines).
    (a)~Plots in the large $N_\text{sat}$ limit (here $N_\text{sat}=100$), showing a good agreement with the power law regime of Eq.~\eqref{eq:omega_power} for two values of $\bar{N}$ and for constant $p_\text{off} = 0.18$ corresponding to $\gamma\simeq 0.116$ and $\gamma\simeq 0.0583$.
    (b)~Plots for a smaller value of $N_\text{sat}$ ($N_\text{sat} = 30$) showing the three distinct asymptotic regimes. Here $\bar{N} = 10 \text{ and } p_\text{off}=0.18 \Rightarrow \gamma\simeq 0.0583$.
    (c)~Plot of the three distinct asymptotic regimes for a higher value of $\gamma$ ($\bar{N} = 10$ and $p_\text{off} = 0.935 \Rightarrow\gamma = 1.49$). The marker at $\omega\tau_1=\tau_1/\tau_{N_\text{sat}}$ denotes the expected position of the low-frequency crossover, while the high-frequency crossover is expected for $\omega\tau_1\approx 1$.}
    \label{fig:supp:oscillatory}
\end{figure}

Finally, at low frequencies $\omega\ll\tau_{N_\text{sat}}^{-1}$, the system again goes to a Maxwell-like rheology:
\begin{subequations}\label{eq:omega_low}
\begin{align}
    \tilde{G}'(\omega) & \underset{\omega\ll\tau_{N_\text{sat}}^{-1}}{\sim} A(\gamma,\bar{N})(\omega\tau_1)^2
    \\
    \tilde{G}''(\omega) & \underset{\omega\ll\tau_{N_\text{sat}}^{-1}}{\sim} 
    B(\gamma,\bar{N})(\omega\tau_1),
\end{align}
\end{subequations}
where the functions $A$ and $B$ take simple forms in the $N_\text{sat}\gg\bar{N}$ limit:
\begin{subequations}
\begin{align}
    A(\gamma,\bar{N}) & =
    \begin{cases}
    (1-e^{1/\bar{N}}) \exp\left[-\frac{N_\text{sat}(\gamma-2)+\gamma+2}{\gamma \bar{N}}\right] \Phi(e^{(2/\gamma-1)/\bar{N}}, 2 , -N_\text{sat}) & \text{if }\gamma<2\\
    \left(1-e^{-1/\bar{N}}\right)e^{(1-2/\gamma)/\bar{N}}\text{Li}_2\left[e^{(2/\gamma-1)/\bar{N}}\right] & \text{if }\gamma>2\\
    \end{cases}\\
    B(\gamma,\bar{N}) & =
    \begin{cases}
    \frac{1-e^{-1/\bar{N}}}{N_\text{sat}}\frac{\exp\left[\left({1}/{\gamma}-1\right)N_\text{sat}/\bar{N}\right]}{\exp\left[\left({1}/{\gamma}-1\right)/\bar{N}\right]-1} & \text{if }\gamma<1\\
    \left(1-e^{-1/\bar{N}}\right)e^{(1-1/\gamma)/\bar{N}}\ln\left[\frac{1}{1-\exp\left[(1/\gamma-1)/\bar{N}\right]}\right] & \text{if }\gamma>1\\
    \end{cases}.
\end{align}
\end{subequations}
Here $\Phi$ denotes the Lerch zeta function defined  as $\Phi(z,s,\alpha) = \sum_{n=0}^{\infty} z^n/(n+\alpha)^s$, which simplifies for $z\ll1$ and $\alpha\gg 1$ (i.e. $p_\text{off} \ll e^{-2}$ and $N_\text{sat}\gg 1$) into $\Phi(z,s,\alpha) \sim \alpha^{-s}/(1-z)$.
$\text{Li}_2$ denotes the polylogarithm function of order 2, which is defined as $\text{Li}_2(x)=\sum_{k=1}^\infty x^k/k^2$. The three successive regimes described by Eqs.~(\ref{eq:omega_high}-\ref{eq:omega_low}) are shown in Fig.~\ref{fig:supp:oscillatory}.

\bibliographystyle{unsrt}
\bibliography{valency_controls_non_exponential_relax}